\documentclass[%
 reprint,
nofootinbib,
 amsmath,amssymb,
 aps,
]{revtex4-1}

\usepackage{graphicx}% Include figure files
\usepackage{dcolumn}% Align table columns on decimal point
\usepackage{bm}% bold math
\begin{document}

\title{HEATING OF A MOLECULAR CLOUD BY A PRIMORDIAL BLACK HOLE}

\author{A.N. Melikhov}
 \email{melikhov94@inbox.ru}
\affiliation{%
Astro Space Center of P.N. Lebedev Physical Institute, Russian Academy of Sciences, Moscow 117997, Russia}%

\author{E.V. Mikheeva}
 \email{helen@asc.rssi.ru}
\affiliation{
Astro Space Center of P.N. Lebedev Physical Institute, Russian Academy of Sciences, Moscow 117997, Russia}%

\date{\today}% It is always \today, today,

\begin{abstract}
The heating of a molecular cloud by photons emitted by a primary black hole (PBH) located inside
the cloud is considered. For graphite and silicate dust particles, the dependence of dust temperature on the distance from PBH is derived, along with the emission spectrum of dust particles. The obtained spectrum is compared with the sensitivity of the Millimetron Space Observatory for various values of concentration and size of dust particles and different PBH masses.
\end{abstract}

\maketitle

%\tableofcontents
\noindent {\small JETP \textbf{166}, 599-611 (2024)}

\noindent {\small DOI: 10.31857/S004445102411e051}

%%%%%%%%%%
\section{Introduction}
%%%%%%%%%%

Interstellar dust is one of the components of the interstellar medium (ISM). It is present in many
astronomical objects, such as the Solar System \cite{hoppe10}, comets and meteoroids \cite{kuppers05}, stellar atmospheres \cite{harvey12}, young stellar objects \cite{keller08}, protostellar and protoplanetary disks \cite{watson09}, reflection nebulae \cite{castellanos11}, supernova remnants \cite{rho09}, molecular clouds \cite{martel12, rice16}, ISM clouds \cite{zhukovska}, galaxies \cite{dunne11}, and active galactic nuclei \cite{haas}, including those at high redshifts \cite{dwek07}.

Analysis of observational data can provide information about both the radiation source and the medium between the object and the observer. It is currently known that dust is the main source of opacity in molecular clouds and star-forming regions, and also serves as material for the formation of
protoplanetary disks and planets. It absorbs radiation, heats up, and re-emits thermal energy in the  infrared range, thereby converting most of the ultraviolet radiation from stars into its own infrared radiation.

Interstellar dust is a complex substance and varies in size, physical structure, and chemical composition
of individual dust particles. The most specific information about dust composition can be obtained
from observing individual spectral features such as broad absorption lines (bands) or dust emission in
the infrared range. Absorption details in infrared spectra near certain wavelengths are associated with
the excitation of vibrational degrees of freedom of individual interatomic bonds in molecules that make
up dust particles \cite{bochkarev}.

There is a number of evidence indicating that interstellar dust consists of silicate and carbon dust
particles, as well as polycyclic aromatic hydrocarbons (PAH).

The presence of silicate particles is indicated, for example, by the absorption band near $9.7~\mu \rm m$, detected during the study of interstellar dust along the line of sight toward the Galactic center using the ISO \textit{Infrared Space Observatory} (ISO) \cite{kemper}, as well as a strong absorption band near $18~\mu \rm m$ and spectral features caused by ice, which are present in the spectrum of the Becklin-Neugebauer object, a bright infrared source in the molecular cloud OMC-1 \cite{gibb}.

Graphite particles are responsible for excess absorption at a wavelength of about $2175~\rm\AA$ \cite{stecher}. This follows from laboratory experiments with carbon particles, which also showed a strong absorption band near this wavelength \cite{bochkarev}.

The presence of PAH in interstellar dust is indicated by features in the infrared emission
spectra of spiral galaxies near 3.3, 6.2, 7.7, 8.6, 11.3 and 12.7~$\mu\rm m$,  which are associated with vibrational transitions in PAH \cite{cesarsky}.

One of the most successful dust models is the \textit{Mathis-Rumpl-Nordsieck} (MRN) model \cite{mathis831}, in which dust consists of spherical silicate and graphite particles with a power-law size distribution, $n(a) \sim a^{-3.5}$. This model agrees well with the interstellar extinction curve  \cite{draine84}.

Dust particles exchange energy with the environment through the absorption and reemission
of photons, collisions, cosmic ray impacts, and exothermic reactions on their surface, such as
the formation of hydrogen molecules \cite{draine11}. The equilibrium temperature of dust is primarily determined by radiative processes in diffuse regions of the interstellar medium, while in
dense molecular clouds ($n > 10^5$~cm$^{-3}$) and in the coronary gas with $T \geq 10^5$~K collisions with gas can play an important role in dust particle energetics. Photons can knock electrons out of dust particles through the photoelectric effect, causing them to become positively charged, however, most of the incident energy is dissipated through vibrational modes within the solid body, i.e., the energy converts to dust heating, which is balanced by cooling through its thermal radiation. Therefore, for dust particles, like for any thermodynamic equilibrium systems, Kirchhoff's law is satisfied \cite{bochkarev, draine11, tielens}.

The absorption characteristic of interstellar dust is represented by absorption efficiency:
%(1)
\begin{equation}
\label{eq1}
    Q(\lambda)=\frac{C_{abs}}{{\sigma}_d},
\end{equation}
where $C_{abs}$ is the absorption coefficient, $\sigma_d = \pi a^2$ is the geometric cross-section of the dust particle, $a$ is the dust particle radius. This characteristic shows what fraction of radiation the dust particle absorbs, and on the scale of the entire molecular cloud -- the fraction of radiation that the molecular cloud absorbs.

In the Galaxy, dust is distributed extremely inhomogeneously and has a hierarchical structure (``clusters within clusters''). The scale of inhomogeneities varies from $10^{-4}$~pc (size of the Solar system) to $10^3$~pc (size of spiral arms). Molecular clouds represent peaks in density distribution at scales corresponding to observed concentrations of interstellar gas and dust \cite{elmegreen}. Interstellar clouds of the same size and mass can have completely different morphological structures.

Dust in the interstellar medium is heated predominantly by ultraviolet radiation from stars. The problem of radiation transfer in molecular clouds is currently well studied. However, dust heating by objects such as primordial black holes (PBHs), which have been receiving increasing attention lately, has not been considered yet.

PBHs are black holes (BHs) that could have formed in the early stages of the Universe's evolution.
The fundamental possibility of their formation was first considered more than 50 years ago \cite{zeldovich67,hawking71}, and interest in them is growing. The main mechanism of PBH formation is the gravitational collapse of matter inhomogeneities if their density contrast  $\delta\equiv\delta\rho/\rho\sim1$ \cite{zeldovich67,hawking71}. Among specific realizations of this process are, for example, PBH formation during inflation \cite{clesse,inomata,bellido,ezquiaga}, during the matter-dominated stage \cite{khlopov80,polnarev85,green,harada,carr17}, such processes were also considered in modified gravity theories \cite{barrow96}, during phase transitions \cite{crawford,kodama,jedamzik}, as a result of domain wall collapse \cite{rubin,garriga,deng,deng17,dokuchaev,kusenko,liu}, due to bubble collisions \cite{hawking82,moss,kitajima}, and others. The approximate relationship between
the PBH mass and its formation time appears as follows:
%(2)
\begin{equation}
\label{eq2}
M\sim \frac{c^3t}{G} \simeq 5 \times 10^{-19}\Big(\frac{t}{10^{-23}s}\Big)M_{\odot},  
\end{equation}
where $c$ is the speed of light, $G$ is the gravitational constant, and $M_{\odot}$ is the solar mass. 
This expression implies that if a PBH formed at Planck time ($\sim$10$^{-43}$~s), it would have Planck mass ($\sim$10$^{-5}$~g). If a PBH formed at  $t = 1$~s, then its mass $M = 10^5 M_{\odot}$ is comparable to the masses of supermassive black holes.

The detection of gravitational waves from merging BHs by \textit{Laser Interferometer 
Gravitational-wave Observatory} (LIGO) \cite{abbott16} increased interest in PBHs, leading to a significant rise in the number of papers on this topic. Currently, LIGO has registered about 100 events accompanied by gravitational wave emission \cite{abbott21}.
During data analysis, it was discovered that, firstly, the intrinsic spin of the merging BHs turned out
to be close to zero, which is difficult to explain for astrophysical BHs but quite logical for PBHs, and
secondly, the masses of these BHs are significantly larger than what was expected for objects representing the final stage of massive star evolution (see, for example, review \cite{remillard06} and analysis in \cite{hutsi21}).

The cosmological properties of PBHs are similar to the cosmological properties of cold dark matter,
making them a natural candidate for dark matter (DM). This idea was first proposed in early studies
of PBH \cite{chapline} and was subsequently discussed many times in literature (for example, \cite{clesse17, ivanov94, lacki,  belotsky, frampton, kashlinsky, clesse18, espinosa}). It should be noted that the question of what constitutes the physical carrier of DM remains unresolved. Years of attempts to detect DM particles have not yet yielded positive results \cite{aleksandrovetal21}. Sterile neutrinos with a mass close to 3~eV, for which encouraging results were obtained \cite{serebrov21, barinov22}, cannot fully solve the DM problem. Thus, the absence of an experimentally confirmed DM particle candidate strengthens the status of other options, such as PBHs.

In \cite{bekenstein}, the laws of thermodynamics were applied to BHs. In particular, it was established that the surface area of a BH and surface gravity are analogous to entropy and temperature, respectively. Additionally, BH can be source of thermal radiation \cite{hawking75}. The radiation temperature $T$ is determined by the expression:
%(3)
\begin{equation}
\label{eq3}
 T = \frac{\hslash c^3}{8\pi k_B GM},  
\end{equation}
where $\hslash$ is Planck's constant, $k_B$ is Boltzmann's constant, $M$ is the BH mass. 

The photon flux from the evaporation of one BH is determined by the expression:
%(4)
\begin{equation}
\label{eq4}
    \frac{dN_{\gamma}}{dt\,dE} = \frac{\Gamma}{2 \pi\hslash}\Big[\exp\Big(E/k_BT\Big)-1\Big]^{-1},
\end{equation}
where $E$ is the photon energy, $\Gamma$ is the “grey” factor, which at $GME/\hslash c^3 \ll 1$ depending on the spin of emitted particles $s$ is expressed as follows \cite{macgibbon90}:
\begin{equation}
\label{eq5}
\Gamma \sim
\begin{cases}
16 G^2 M^2 E^2/\hslash^2c^6, \quad s = 0,\\
2 G^2 M^2 E^2/\hslash^2c^6, \quad s = 1/2,\\
64 G^4 M^4 E^4/3\hslash^4c^{12}, \quad s = 1,\\
256 G^6 M^6 E^6/45\hslash^6c^{18}, \quad s = 2.
\end{cases}
\end{equation}
At $GME/\hslash c^3 \gg 1$ the expression for $\Gamma$ takes the form
\begin{equation}
\label{eq6}
 \Gamma \sim \frac{27G^2M^2E^2}{\hslash^2 c^6},  
\end{equation}
which corresponds to black body radiation \cite{macgibbon90}. Figure~1 illustrates the difference between the photon flux ($s = 1$), emitted by PBH with mass $M = 10^{13}$~g, from black body radiation, characterized by flux with $\Gamma$ according to (\ref{eq6}).

%===============================fig1
\begin{figure}[]
 \centering
 \includegraphics[width=0.9\textwidth]{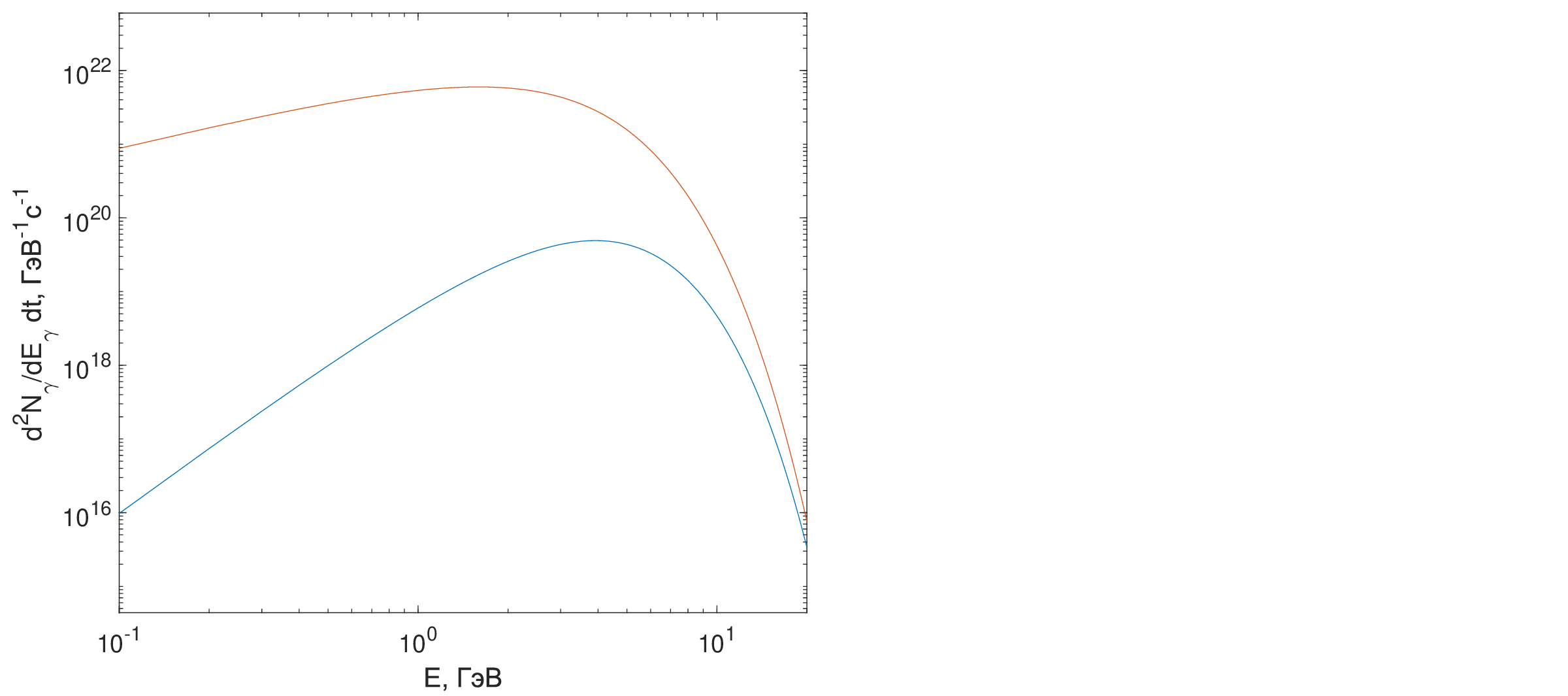}
\caption{%\small 
Photon flux emitted by PBH with mass $M = 10^{13}$~g (blue line), depending on their energy $E$ to black body radiation (red line).}
\end{figure}
%===============================

As seen from (\ref{eq3}) and (\ref{eq4}), high radiation intensity is characteristic of BHs with small masses. However, despite the fact that PBH evaporation ends in a powerful burst of high energy radiation, during which particle energies can reach several hundred TeV, there is currently no experimental confirmation of PBH existence.

Currently, PBHs with initial masses less than $\sim10^{15}$~g have already evaporated, but the consequences of their existence can be detected through their influence on various cosmological processes, such as primordial nucleosynthesis \cite{carr10} and baryogenesis \cite{turner79, barrow91, upadhyay99, bugaev03}. Evaporating PBHs can also be additional sources of neutrinos \cite{bugaev03, bugaev02}, gravitinos \cite{khlopov06}, and other particles \cite{green99, lemoine00}.

Evaporating PBHs can contribute to galactic \cite{lehoucq} and extragalactic gamma-ray backgrounds \cite{page}; they are invoked to explain antimatter particles in cosmic rays \cite{barrau}, such PBHs can also be sources of annihilation line radiation from the Galactic center \cite{okele} and cause some
short period gamma-ray bursts \cite{cline}.

PBHs of larger masses can influence the large scale structure of the Universe \cite{meszaros275, carr77, freese83, carr83, afshordi03}, be seeds of supermassive BHs \cite{carr84, duchting04, khlopov05, bean02}, be \textit{Massive Compact Halo Objects} (MACHOs), and one of the
sources of X-ray background due to accretion of matter onto PBHs \cite{ricotti08}.

To date, there is no convincing evidence for the existence of PBHs. Nevertheless, the study of the aforementioned effects allows to impose constraints on the number of 
PBHs\footnote{Constraints on the number of PBHs are determined using the quantity $f\equiv\Omega_{PBH}/\Omega_{DM}$, i.e., for the relative contribution of PBHs to DM.} 
and, thereby, on the cosmological models that produce them. Currently, constraints have been found on the cosmological density of PBHs, both those that have already evaporated at present and fairly large ones up to $M\sim 10^{50}$~g. Various physical effects were involved in finding the constraints. There are distinctions between constraints due to evaporation of black holes (which can contribute to the extragalactic gamma-ray background, positron flow, and annihilation line from the Galactic center), lensing, gravitational waves, various dynamical effects (such as wide binary systems, stellar clusters, halo dynamical friction, etc.), accretion, distortions of cosmic microwave background
and large scale structure (arising from the requirement that various cosmological structures not form earlier than observed), for more details see review \cite{carr21}. Also, constraints on the number of PBHs based on the interaction of Hawking radiation components with ISM were obtained in papers \cite{kim, melikhov22, melikhov23}.

This paper examines the heating of dust by photons emitted from PBHs due to their evaporation in a spherically symmetric molecular cloud, and calculates the dependence of dust temperature on the distance to PBHs and the emission spectrum of such a cloud.

%%%%%%%%%%%%%%%%%%%%%%%%%%%%%%%%%%%%%%%%%%%%%%%%
\section{DETERMINATION OF DUST TEMPERATURE AS A FUNCTION OF DISTANCE TO PBH}
%%%%%%%%%%%%%%%%%%%%%%%%%%%%%%%%%%%%%%%%%%%%%%%%

The largest number of PBHs can be expected in the central part of the Galaxy due to its higher density. Therefore, we will consider molecular clouds located in the Central Molecular Zone of the Galaxy (CMZ) as a reservoir for PBHs. Let's consider the mass range of PBHs from $10^{16}$ to $10^{20}$~g.

Let's assume that the molecular cloud is spherically symmetric with a radius of 5~pc. At the center of the cloud, there is a PBH with mass $M$, which emits X-ray and gamma photons according to eq.~(\ref{eq4}). The thermal structure of the cloud is determined by the interstellar dust inside it, which absorbs, scatters, and re-emits in a continuous spectrum. It is also assumed that the cloud consists of dust grains of the same size, and their temperature depends on the radius (distance to PBH). Since X-ray and gamma radiation weakly interacts with interstellar 
dust\footnote{
Interstellar dust most effectively absorbs radiation in the visible and ultraviolet ranges \cite{draine84} while PBHs have high intensity in X-ray and gamma-ray ranges (with photon energies 1~keV$-100$~MeV). Dust grains absorb such radiation weakly. Dust absorption of photons in this energy range has been poorly studied. Currently, there are works that examine dust absorption of photons with energies up to 10~keV (see, for example, \cite{corrales}), or up to 1~MeV \cite{dwek96}. Photons with such energies are emitted by PBHs with masses up to $\sim 10^{18}$~g and up to $\sim 10^{16}$~g, respectively.}, we will only consider absorption, neglecting scattering. Moreover, the scattering efficiency for the considered energy range is still unknown.

The spectral luminosity of one dust grain equals
%(7)
\begin{equation}
\label{eq8}
 L_{gr} = 4\pi a^2 Q(\lambda) B_{\lambda}(T_d),
\end{equation}
where $B_{\lambda}(T_d)$ is the Planck function. It is also assumed that the absorption of energy by dust is continuous, i.e., the corpuscular properties of radiation are neglected. In this case, the dust
temperature is found from the condition of radiative equilibrium:
%(8)
\begin{equation}
\label{eq9}
\pi a^2 \int u_{\lambda}\,c\,Q(\lambda)\,d\lambda = \int L_{gr}\,d\lambda,
\end{equation}
where $u_{\lambda}$ is the radiation field energy density. 

This equation can be rewritten as follows:
%(9)
\begin{equation}
\label{eq10}
\frac{L_{PBH}(M)}{4\pi r^2} = 4\pi \int Q(\lambda) \, B_{\lambda}(T_d) \,d\lambda,
\end{equation}
where $L_{PBH}(M)$ is the luminosity of PBH with mass $M$, which is determined by the following
expression:
%(10)
\begin{equation}
\label{eq11}
L_{PBH}(M) = \int_0^EQ(E\,)E\,\frac{\Gamma}{2 \pi \hslash}\, \Big(e^{\frac{E}{k_BT}}-1\Big)^{-1}\,dE\,.
\end{equation}
Integration is carried out up to $E = 10^6$~eV (see footnote 2).

Equation (\ref{eq9}) is valid only under the assumption that dust is heated and radiates continuously, and the same equilibrium temperature is established for all dust grains of the same size. This assumption can only be valid for large grains ($a\geq0.01~\mu$m). Meanwhile, it is known that dust grains of different sizes react differently to absorption depending on the ratio of photon energy and the thermal energy of the dust grain \cite{kruegel,duley,draine01}. Small grains have low heat capacity, therefore when absorbing small portions of energy, the dust grain temperature rises sharply. Thus, such particles are characterized by abrupt temperature changes, i.e., the thermal evolution of the dust grain is stochastic. Between temperature jumps, most small particles cool down to the temperature of cosmic microwave background radiation. The dust grain emits when its temperature $T$ is above equilibrium \cite{bochkarev}. The stochastic mode of dust heating and radiation occurs when the photon energy is greater than or comparable to the dust grain energy. Therefore, we will consider relatively large dust grains ($a\geq0.01~\mu$m).

The absorption efficiency for X-ray and gamma photons was taken from  \cite{dwek96} (Table~5.1 for graphite and Table~5.2 for silicate dust grains\footnote{For the constraints on the PBH fraction in DM obtained in \cite{melikhov22, melikhov23}, it is important how the function $Q(\lambda)$ depends on wavelength. If instead of $Q$, considered in \cite{tielens}, we use what was described in \cite{dwek96}, the constraint curve rises, and the constraints disappear.}). Calculations were performed for dust grains with sizes $a = 0.01, 0.02, 0.05,$ and $0.1~\mu\rm m$. For visible and ultraviolet radiation photons, we will use the following approximation \cite{tielens}:
%(11)
\begin{equation}
\label{eq7}
    Q(\lambda) =
    \begin{cases}
    1, &\text{$\lambda \leq 2\pi a$}\\
    \frac{2\pi a}{\lambda}, &\text{$\lambda > 2\pi a$}.
    \end{cases}
\end{equation}

The Galactic Central Molecular Zone contains the highest concentration of high-density molecular
gas \cite{ferriere}. The total mass of gas in the CMZ is $3_{-1}^{+2}\cdot10^7\,M_{\odot}$ \cite{dahmen}, with 10\% of this mass concentrated in the Sgr B2 molecular cloud  \cite{gordon}, located 100~pc from the center of the Milky Way. The remaining molecular gas is dispersed among dozens of giant molecular clouds whose masses are $1-2$~orders of magnitude lower  \cite{longmore}.

Various methods have established that the average gas concentrations in the CMZ molecular clouds
range from $\sim 10^{4}$~cm${}^{-3}$ to $\gtrsim 10^{5}$~cm${}^{-3}$ \cite{bally, tsuboi, jackson, nagai, jones12, jones13, gusten, zylka}. 
Observations in CS lines of some molecular clouds indicate that their highest concentrations can vary from $10^5$~cm${}^{-3}$ to $10^6$~cm${}^{-3}$ \cite{serabyn91, serabyn92}. Analysis of data obtained from observations in HC$_3$N, lines showed that even higher gas concentration can be found in the Sgr~B2 cloud. In this cloud, the average gas concentration is $10^{5}$~cm${}^{-3}$, and in its core it exceeds $10^{7}$~cm${}^{-3}$ \cite{lis}. A similar analysis was presented for the molecular cloud M-0.02-0.07. It showed that it contains regions with lower ($\sim10^{3}$~cm${}^{-3}$) and higher ($\sim10^{5}$~cm${}^{-3}$) concentrations \cite{walmsley}.

Thus, the gas concentration in CMZ molecular clouds ranges from $10^3$~cm${}^{-3}$ to $10^7$~cm${}^{-3}$. Considering that dust mass comprises approximately $1\%$ of gas mass \cite{tielens, bochkarev}, we find that dust concentration varies from $n_d = 10^{-5}$~cm${}^{-3}$ to $n_d = 10^{-1}$~cm$^{-3}$ for clouds consisting of dust grains with size $a=0.01~\mu\rm m$, from $n_d = 10^{-6}$~cm$^{-3}$ to $n_d = 10^{-2}$~cm$^{-3}$ -- with size $a=0.02\mu\rm m$, from $n_d = 10^{-7}$~cm$^{-3}$ to $n_d = 10^{-3}$~cm$^{-3}$ -- with size $a=0.05 \mu\rm m$, and from $n_d = 10^{-8}$~cm$^{-3}$ to $n_d = 10^{-4}$~ cm$^{-3}$ -- with size $a=0.1~\mu\rm m$. These values will be used in further calculations.

We can estimate the optical depth value
%(12)
\begin{equation}
\tau(E) = \int_0^r \pi a^2 n_d\, Q(E)\,dr^\prime,
\end{equation}
for the parameter values described above. Figures~2 and 3 show graphs of optical depth dependence
on energy for graphite and silicate dust grains with different sizes and dust concentrations in the
molecular cloud. Despite the fact that photons emitted by PBH with given masses interact poorly
with dust grains, at the boundary of a 5~pc molecular cloud, the medium becomes optically thick at certain dust concentrations and photon energies.

%====================================fig2
\begin{figure}
\includegraphics[width=0.5\textwidth]{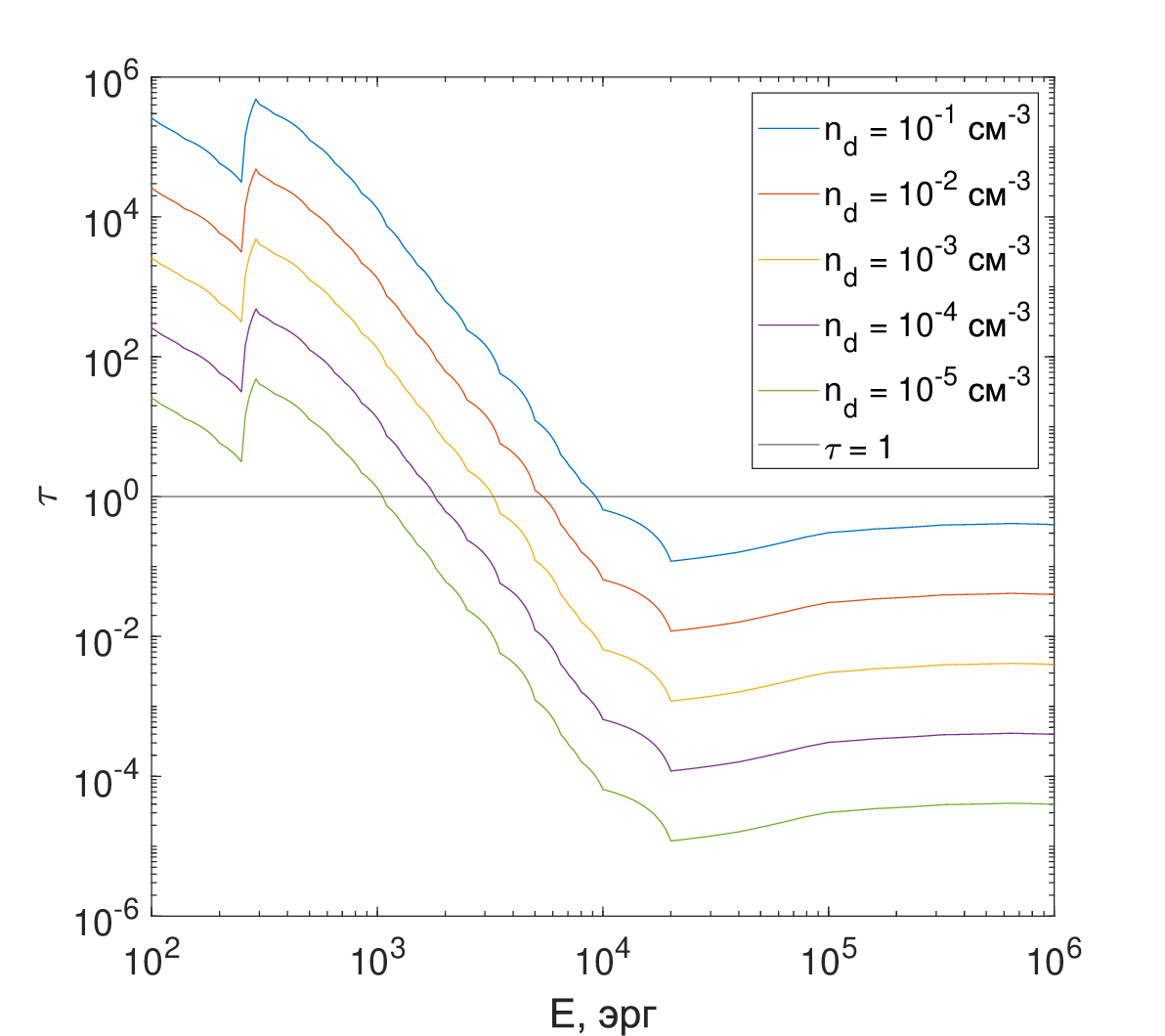}
\includegraphics[width=0.5\textwidth]{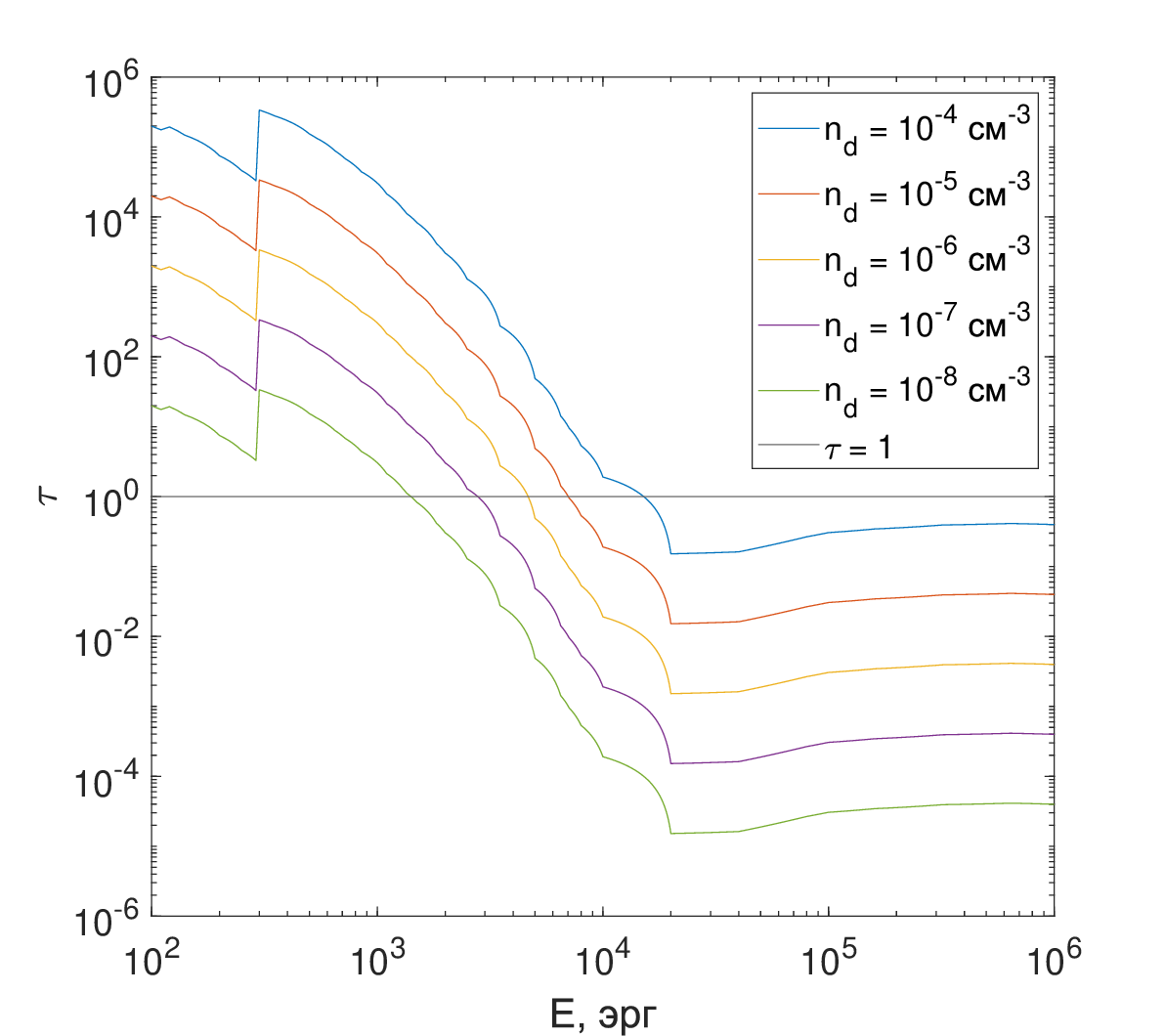}
\caption{
Dependence of optical depth as a function of energy during irradiation of \textit{graphite} dust grains by photons emitted by PBHs. Data are presented for a molecular cloud with a radius of 5~pc, consisting of dust grains with sizes of $a=0.01~\mu$m (\textit{top}) and $a=0.1~\mu$m (\textit{bottom}). Colored lines show dust grain concentrations. The black horizontal line indicates the optical depth $\tau = 1$.}
\label{fig:tau_graph}
\end{figure}

%====================================fig 3
\begin{figure}
\label{fig:tau_sil}
\includegraphics[width=0.5\textwidth]{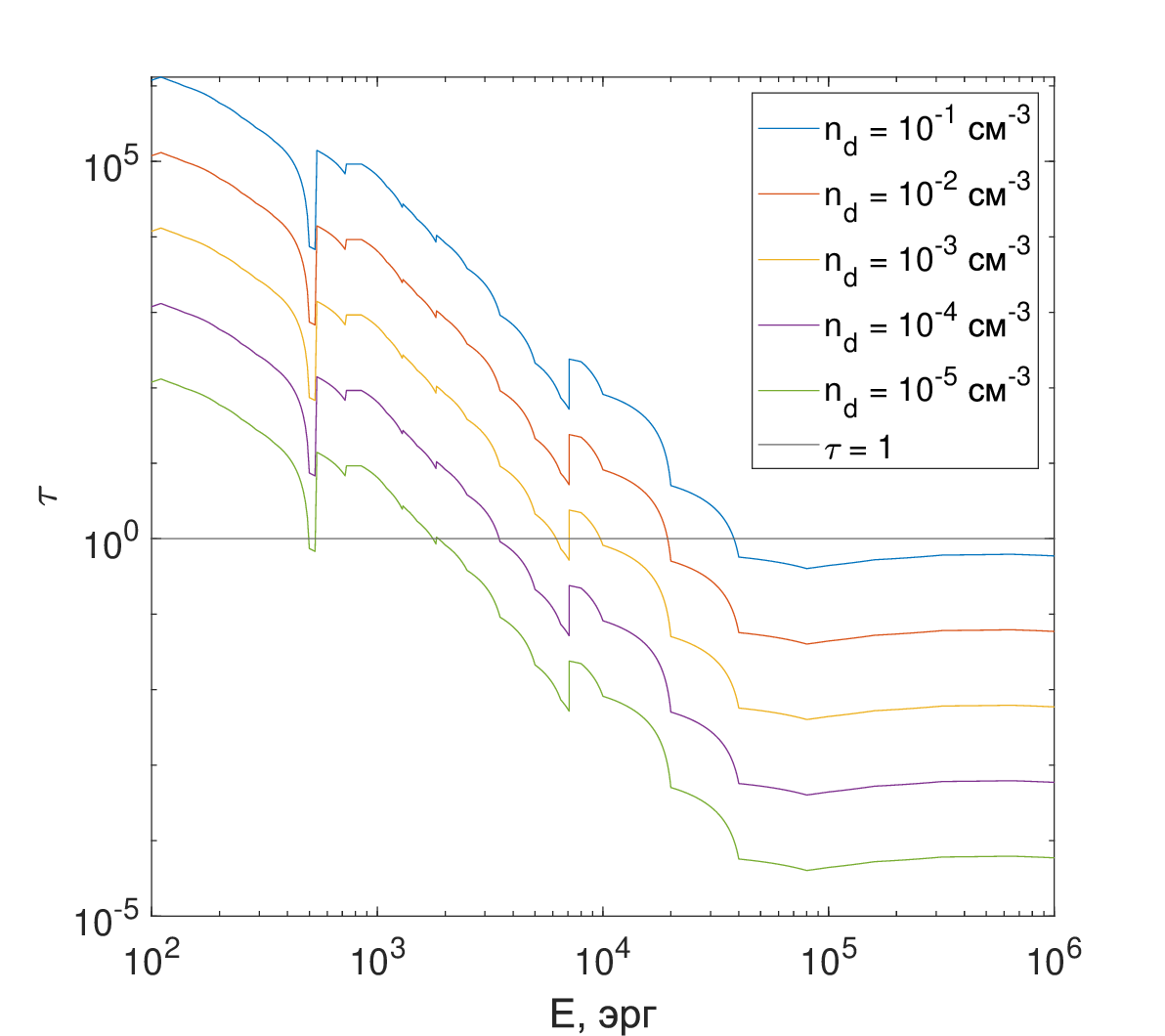}
\includegraphics[width=0.5\textwidth]{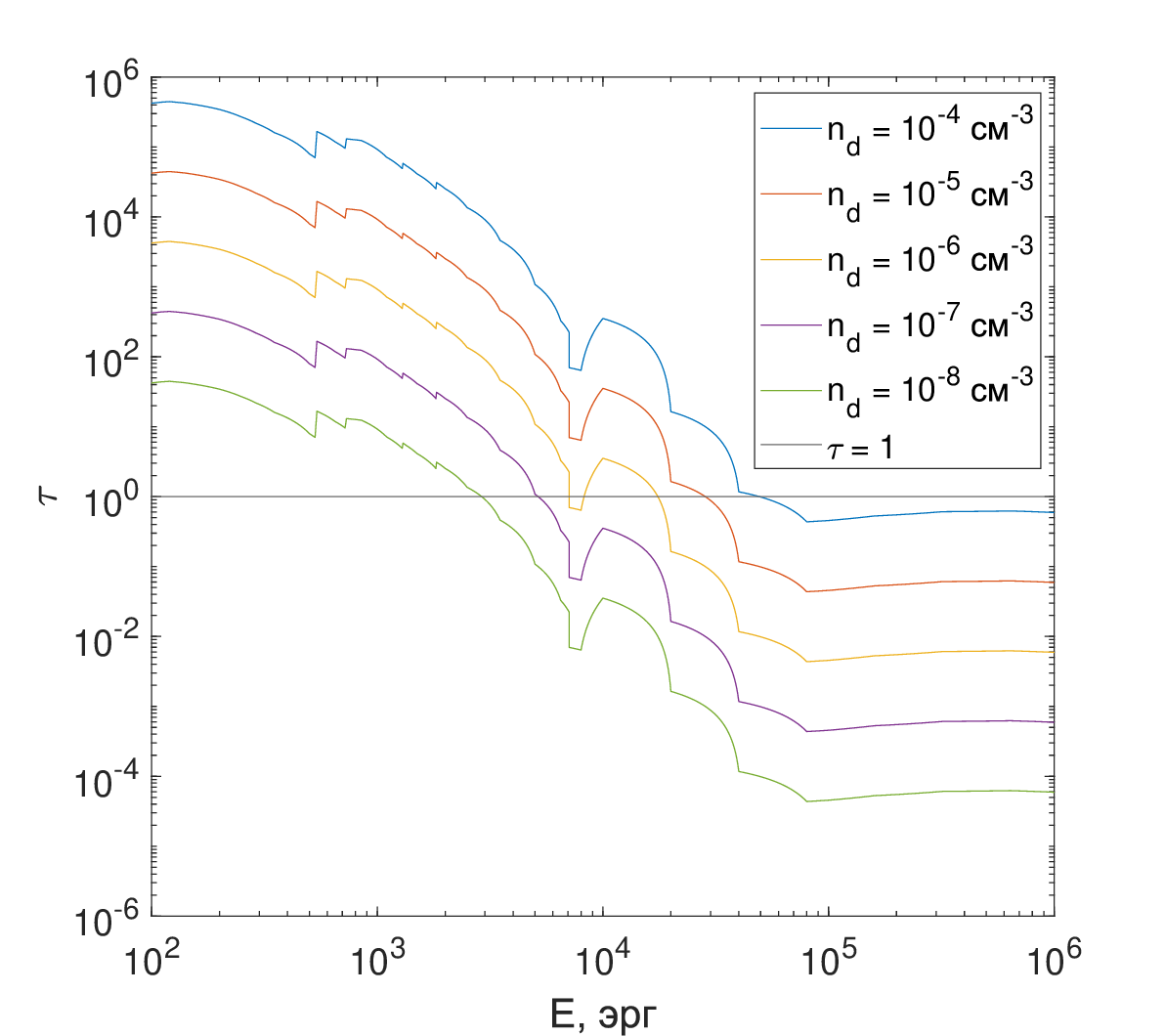}
\caption{
Dependence of optical depth as a function of energy during irradiation of \textit{silicate} dust grains by photons emitted by PBHs. Data are presented for a molecular cloud with a radius of 5~pc, consisting of dust grains with sizes of $a=0.01~\mu$m (\textit{top}) and $a=0.1~\mu$m (\textit{bottom}). Colored lines show dust grain concentrations. The black horizontal line indicates the optical depth $\tau = 1$.}
\end{figure}
%=====================================

From equation (\ref{eq10}) we obtain the dependence of dust grain temperature on distance to PBH:
%(13)
\begin{equation}
T(r) = CL_{PBH}^{1/5}(M)r^{-2/5},
\end{equation}
where $C$ is some constant.

Figure 4 illustrates the temperature dependence of graphite and silicate dust grains on the distance
to PBH with masses of $10^{16}$~g, $10^{18}$~g, and $10^{20}$~g. As shown in the graph, the dust temperature drops sharply with distance to PBH. Dust grains are heated to a temperature of $T_d \geq 3$~K in a spherical layer with a radius of $r \lesssim 10^3$~cm. Since in this case the temperature weakly depends on the dust particle size, we only provide this graph and do not show graphs for dust grains with sizes 0.02, 0.05, and 0.1~$\mu$m.

%===================================fig4
%\onecolumn
\begin{figure}
\label{fig:temp}
\includegraphics[width=0.5\textwidth]{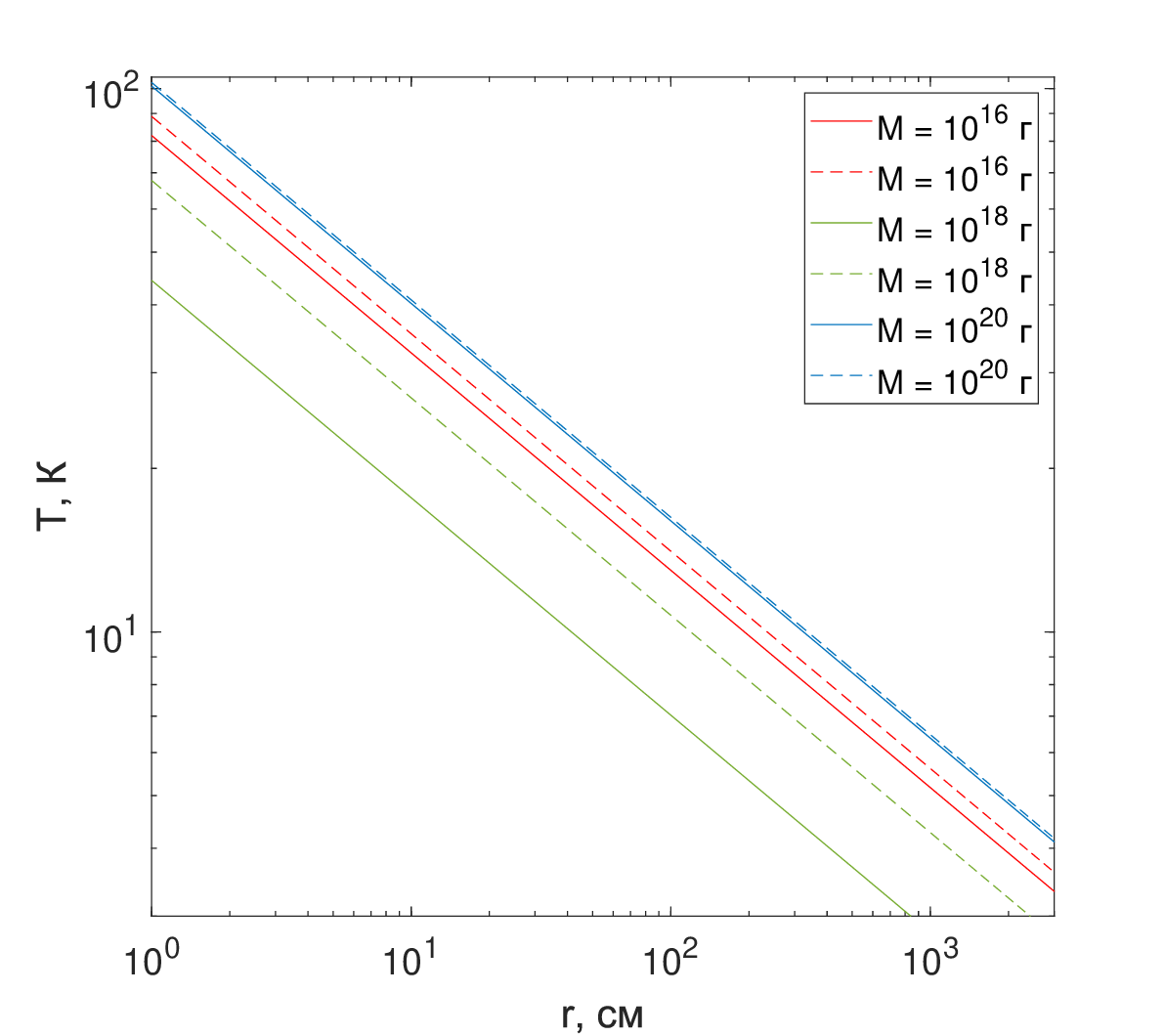}
\caption{
The temperature dependence of \textit{graphite} (solid lines) and \textit{silicate} (dashed lines) dust grains on the distance to PBHs with masses $M=10^{16}$~g, $M=10^{18}$~g, and $M=10^{20}$~g 
(see legend). The graph is shown for dust grains with a size $a=0.01~\mu$m.
}
\end{figure}
%\twocolumn
%===================================

%%%%%%%%%%%%%%%%%%%%%%%%%%%%%%%%
\section{EMISSION SPECTRUM \\OF THE MOLECULAR CLOUD}
%%%%%%%%%%%%%%%%%%%%%%%%%%%%%%%%

The infrared radiation flux from the molecular cloud heated by PBH with mass $M$, is determined
as follows:
%(14)
\begin{equation}
F_{ir}=\int_{0}^{r}n_dL_{gr}dr^{\prime},
\end{equation}
where $r$ is the distance from PBH at which dust heating is most significant. The emission spectra
of graphite and silicate dust grains heated by PBH with different masses, at different concentrations
and dust grains sizes are shown in Figs.~5--9. The sensitivity data from the Millimetron Space Observatory is overlaid with obtained spectra. As can be seen from the figures, there is a fundamental possibility to detect dust heating by Hawking radiation if the dust concentration in the cloud lies in the range from $n_d=10^{-4}$~cm$^{-3}$ to  $n_d=10^{-1}$~cm$^{-3}$ (while the size of individual dust grains in the cloud varies from 0.1 to 0.01~$\mu$m, respectively). In other cases, the spectral curves are below the Millimetron sensitivity curve. This important conclusion is limited by the applicability of the considered model constraints (including the assumption that PBH radiation is the only source of dust heating), but can explain the presence of very compact ``hot'' spots on the temperature distribution map of the molecular cloud (in future observations).

%=================================fig5
\begin{figure*}
\includegraphics[width=0.49\textwidth]{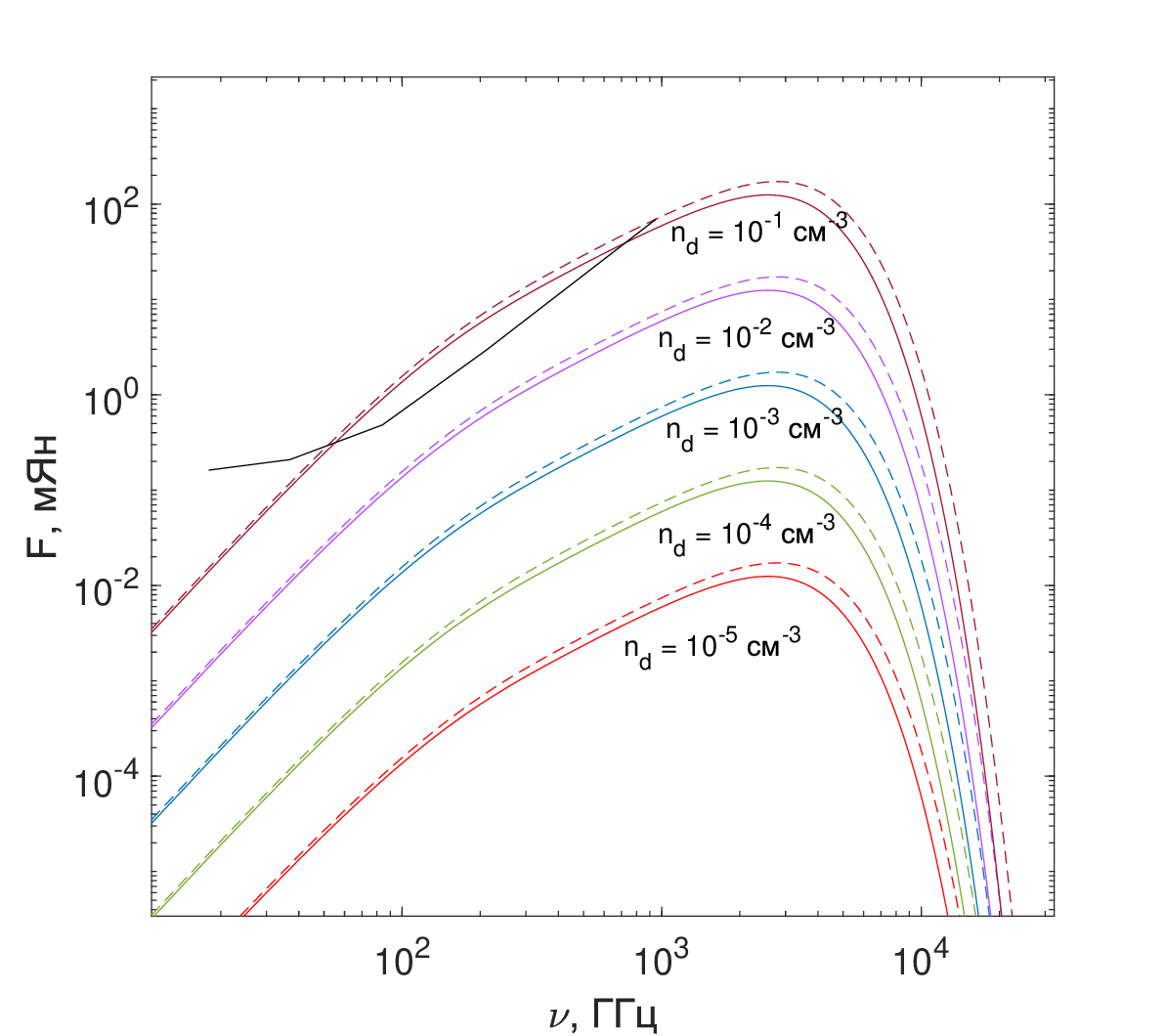}
\includegraphics[width=0.49\textwidth]{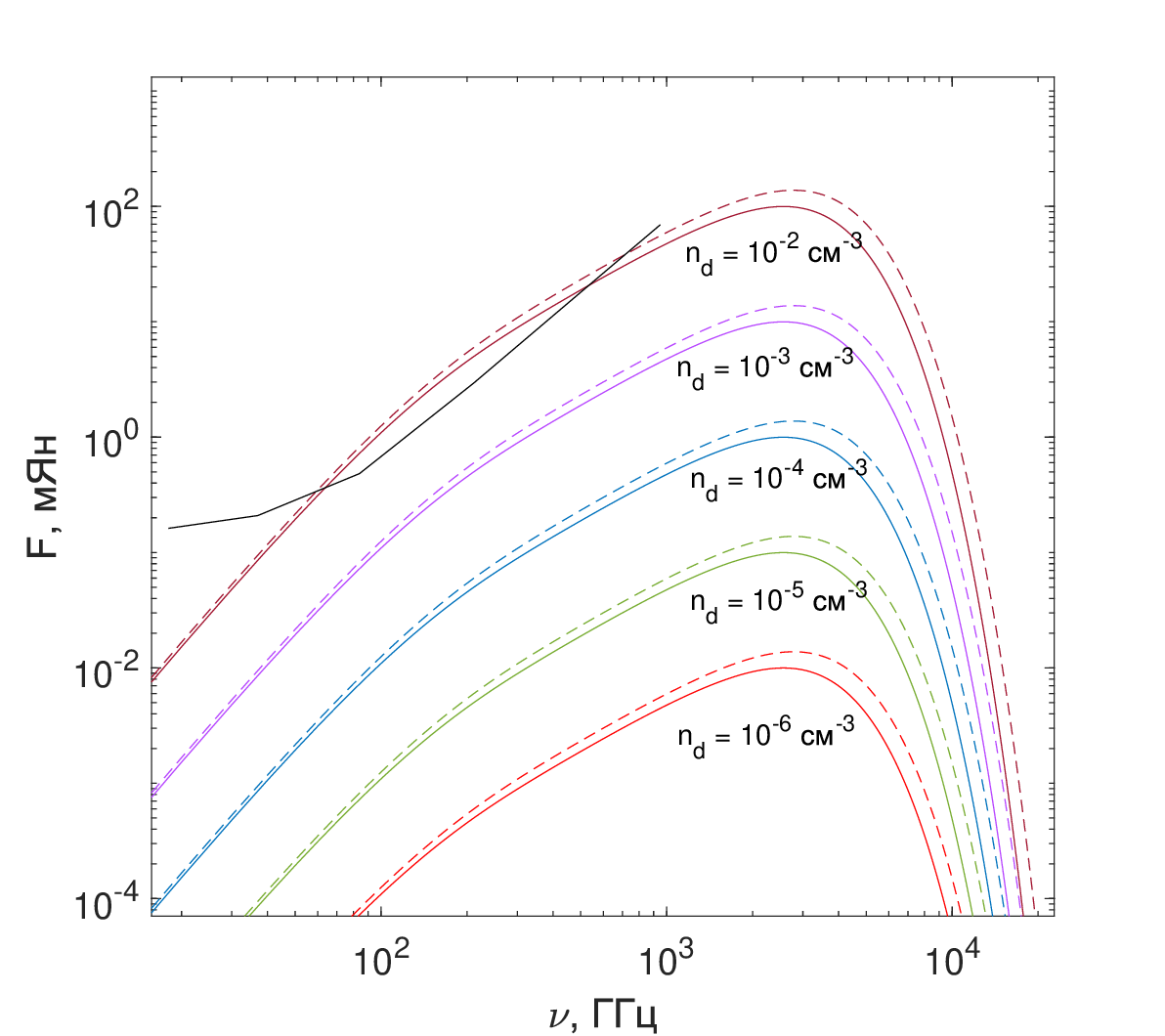}\\
\includegraphics[width=0.49\textwidth]{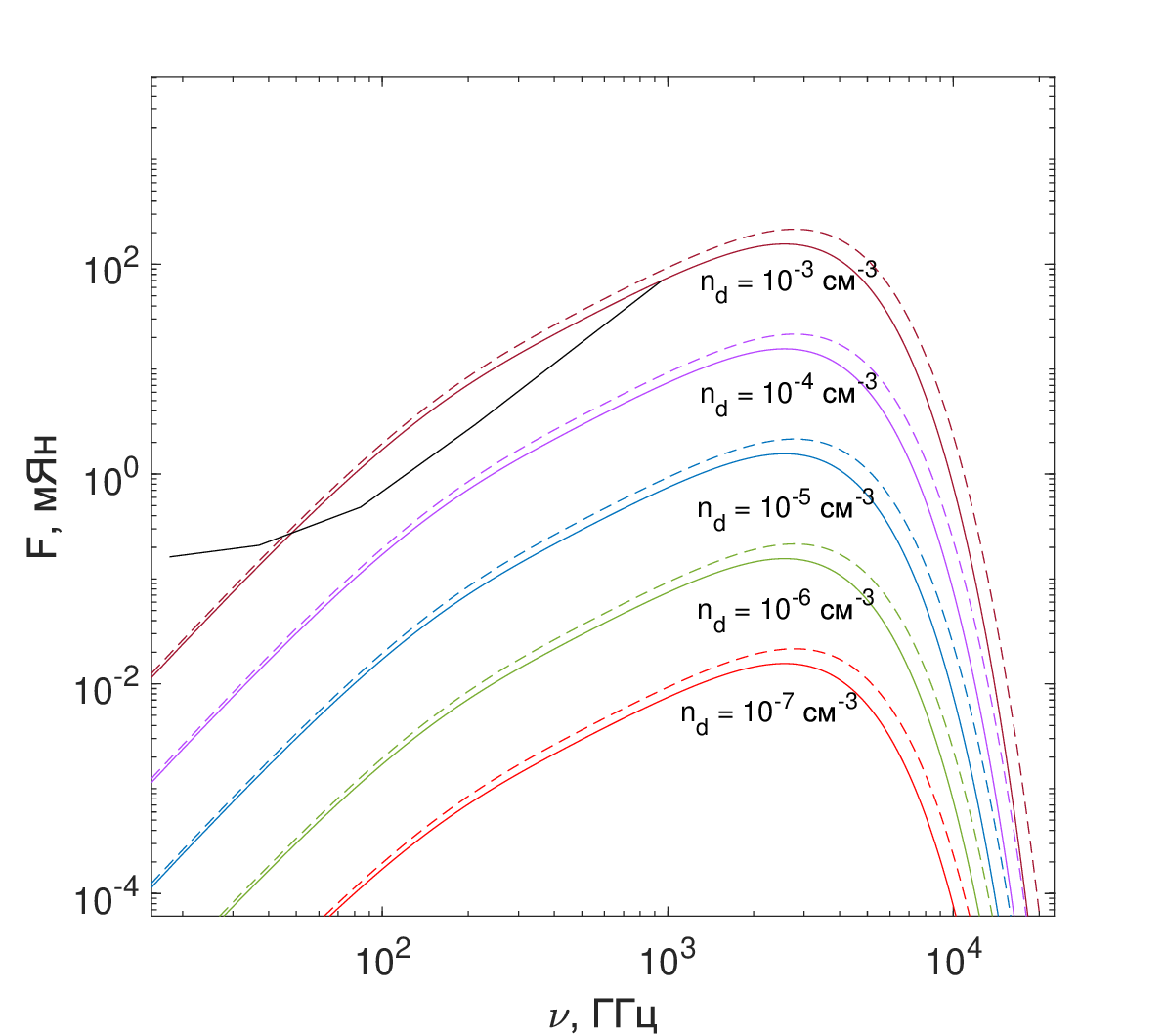}
\includegraphics[width=0.49\textwidth]{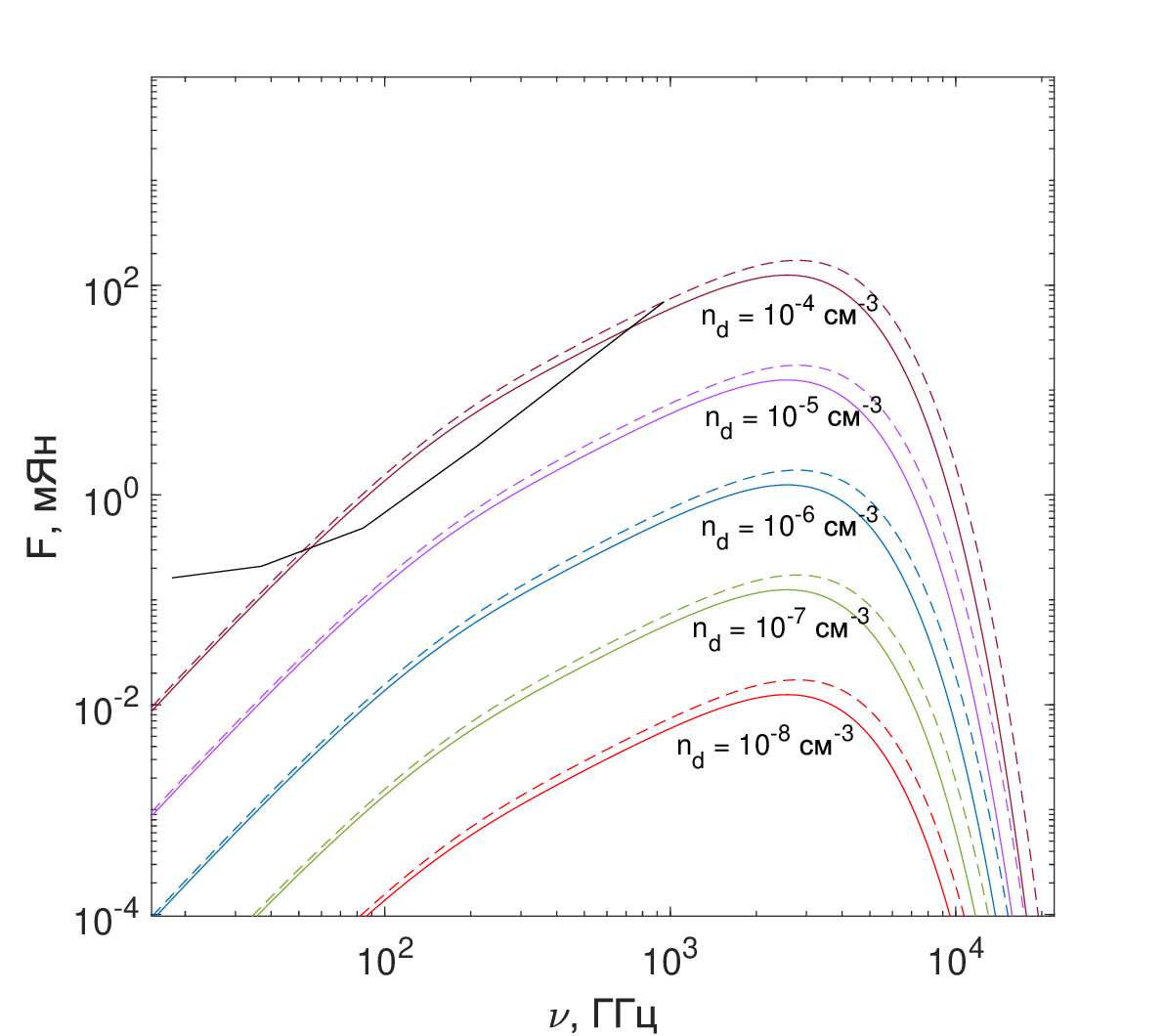}
\caption{
Emission spectra of dust grains heated by PBH with mass $M=10^{16}$~g for different dust concentrations in the molecular cloud. 
Dust grain sizes $a = 0.01~\mu$m (\textit{top left}), $a = 0.02\mu$m (\textit{top right}), $a = 0.05~\mu$m (\textit{bottom left}), $a = 0.1~\mu$m (\textit{bottom right}). 
Solid colored lines represent spectra of grafite grains, dashed colored lines represent silicate grains. 
The black line shows the sensitivity curve of the Millimetron Space Observatory in interferometer mode, built on the Millimetron -- ALMA base using the sensitivity calculator developed at LPI ASC (see millimetron.radioastron.ru). 
The bandwidth of the Millimetron receiver was assumed to be 4~GHz, with an integration time of $15-120$~s.
}
\end{figure*}

The appearance of the spectra and the overlaid sensitivity curve of the planned Millimetron Space Observatory in Figs.~5--9 looks very similar at first glance. However, the spectral fluxes for particles of different sizes (and different concentrations due to the fixed mass fraction of dust in the cloud) differ significantly. Also, the specific form of the spectrum is sensitive to the dust temperature as a function of distance to PBH, while the dust temperature depends on the absorption efficiency as a function of Hawking radiation photon energy.

%===============================fig6
\begin{figure*}
%\label{fig:spec_10_17}
\includegraphics[width=0.49\textwidth]{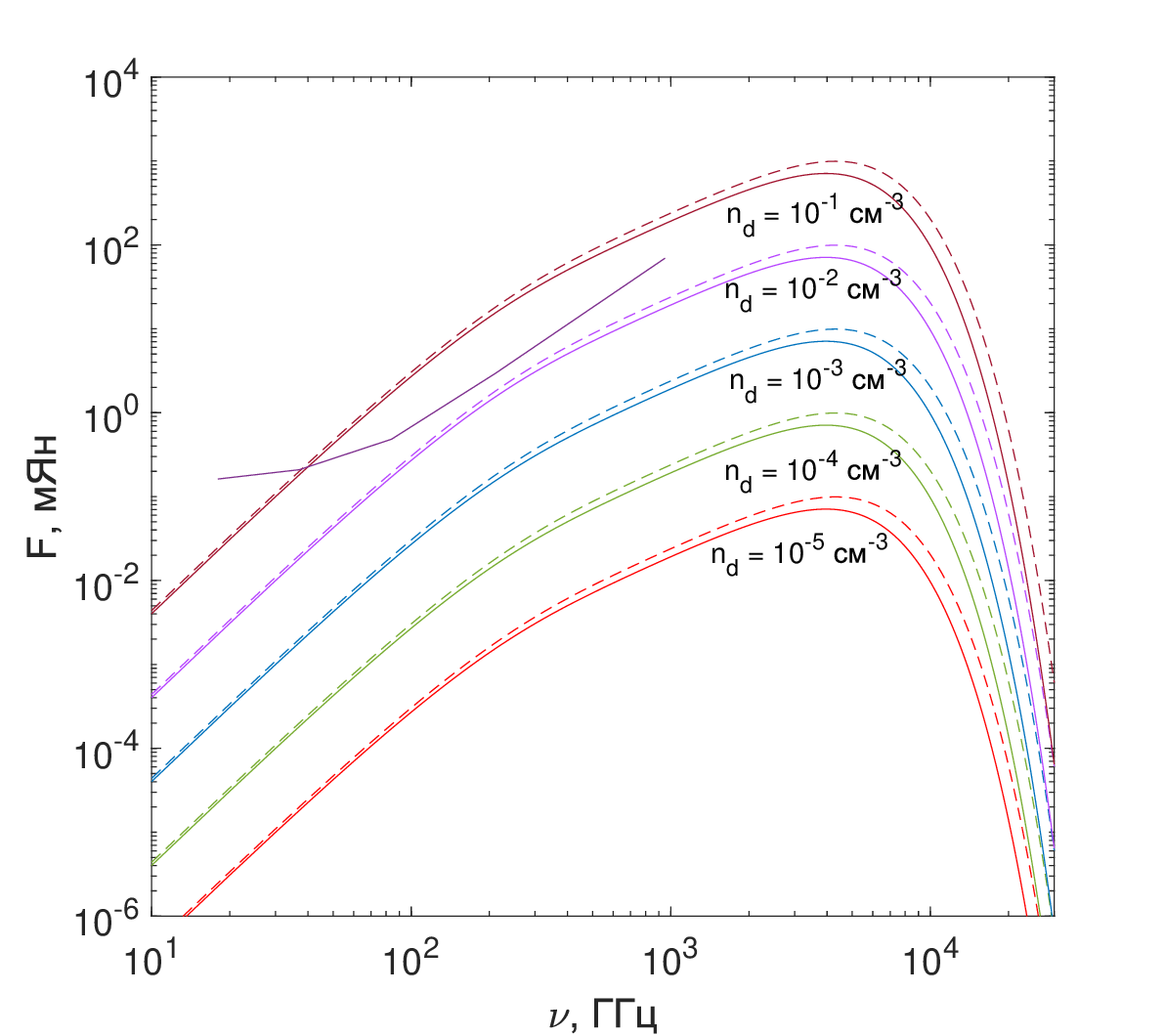}
\includegraphics[width=0.49\textwidth]{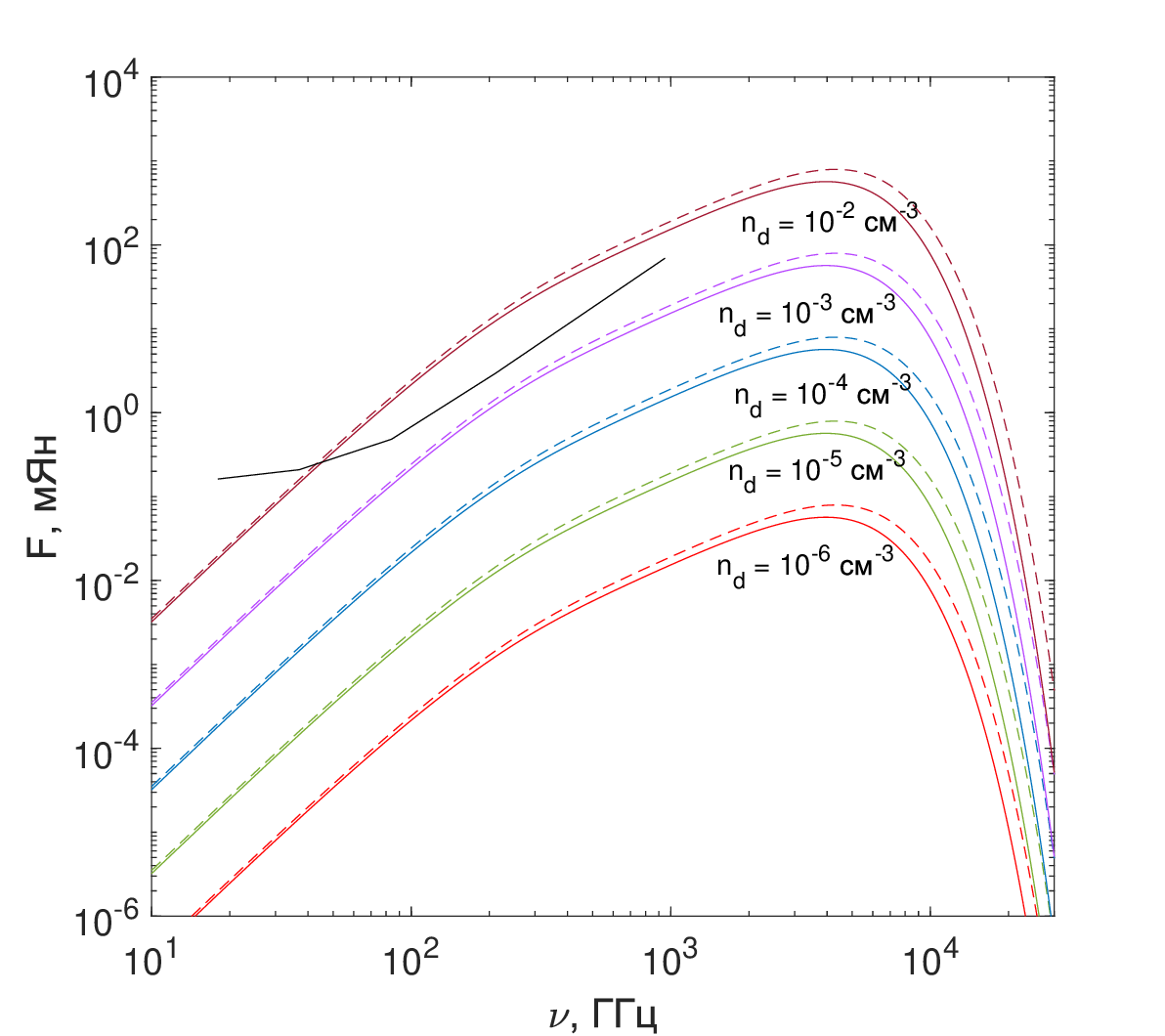}\\
\includegraphics[width=0.49\textwidth]{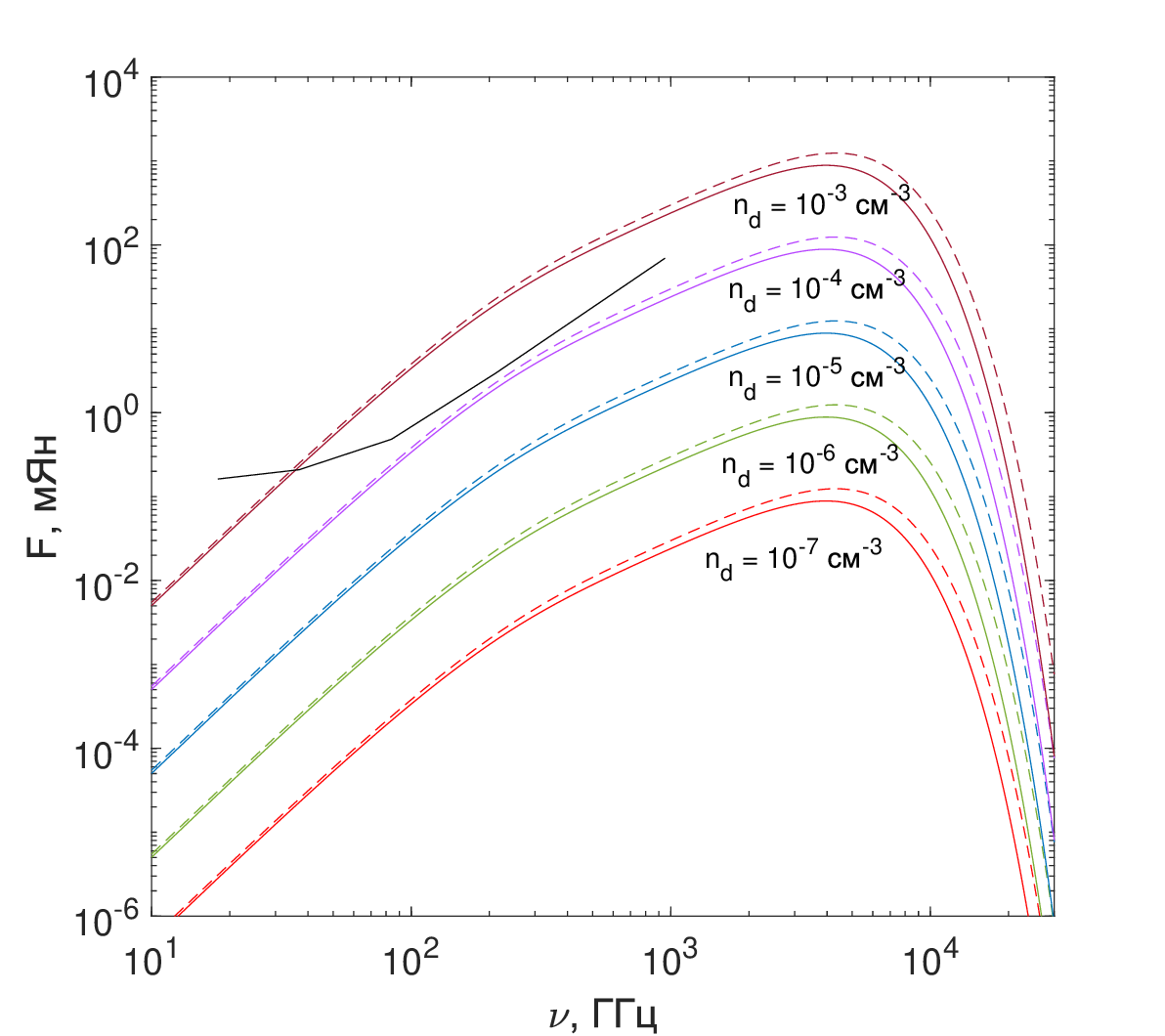}
\includegraphics[width=0.49\textwidth]{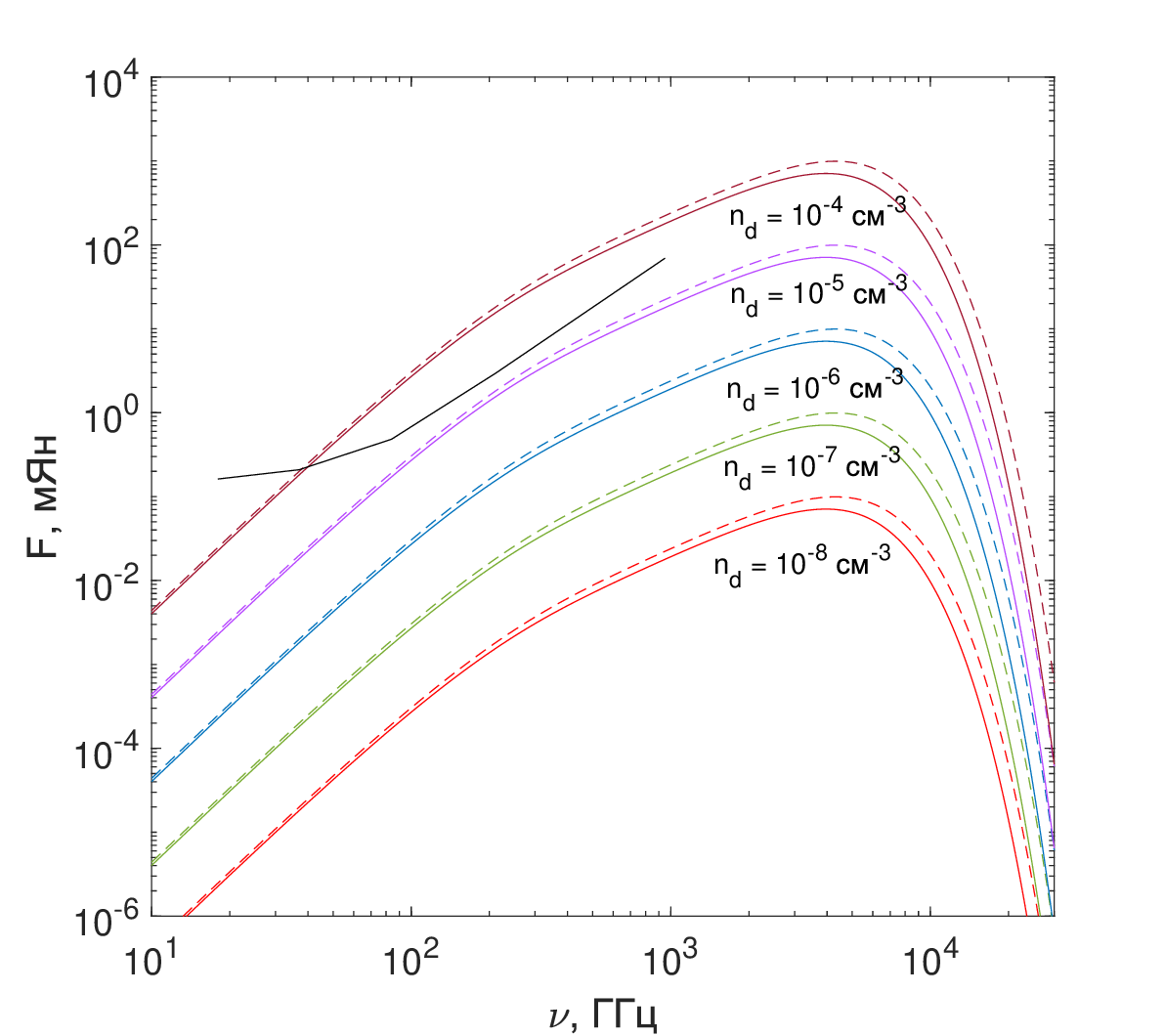}
\caption{Same as in Fig.~5, but for $M=10^{17}$~g.}
\end{figure*}

Note that the accretion of matter onto a PBH will not affect its evolution while inside the molecular
cloud, since its lifetime is severely limited. Thus, in work \cite{jefferson032018}, the lifetime of molecular clouds was calculated based on the large-scale dynamics of the ISM in the Galaxy. In this model, the lifetime of a molecular cloud depends on the distance to the Galactic center and is defined as
%(15)
\begin{equation}
\tau = \vert\tau_{\kappa}^{-1} + \tau_{\Omega_P}^{-1} + \tau_{ff,g}^{-1} + \tau_{cc}^{-1} - \tau_{\beta}^{-1}\vert^{-1}, 
\end{equation}
where $\tau_{\kappa}$ is the contribution of epicyclic perturbations, 
$\tau_{\Omega_P}$ is the contribution of spiral arm crossings of the interstellar medium, 
$\tau_{ff,g}$ is the contribution of interstellar cloud collapse, 
$\tau_{cc}$ is the contribution of cloud-cloud collisions, 
$\tau_{\beta}$ is the contribution of galactic shear, 
i.e., the time scale on which the cloud is torn apart by differential rotation. According to this model, in the Milky Way at distances of $4-10$~kpc from the Galactic center, the lifetime of a molecular cloud ranges from 20 to 60 million years.

%==============================fig7
%\onecolumn
\begin{figure*}
%\label{fig:spec_10_18}
\includegraphics[width=0.49\textwidth]{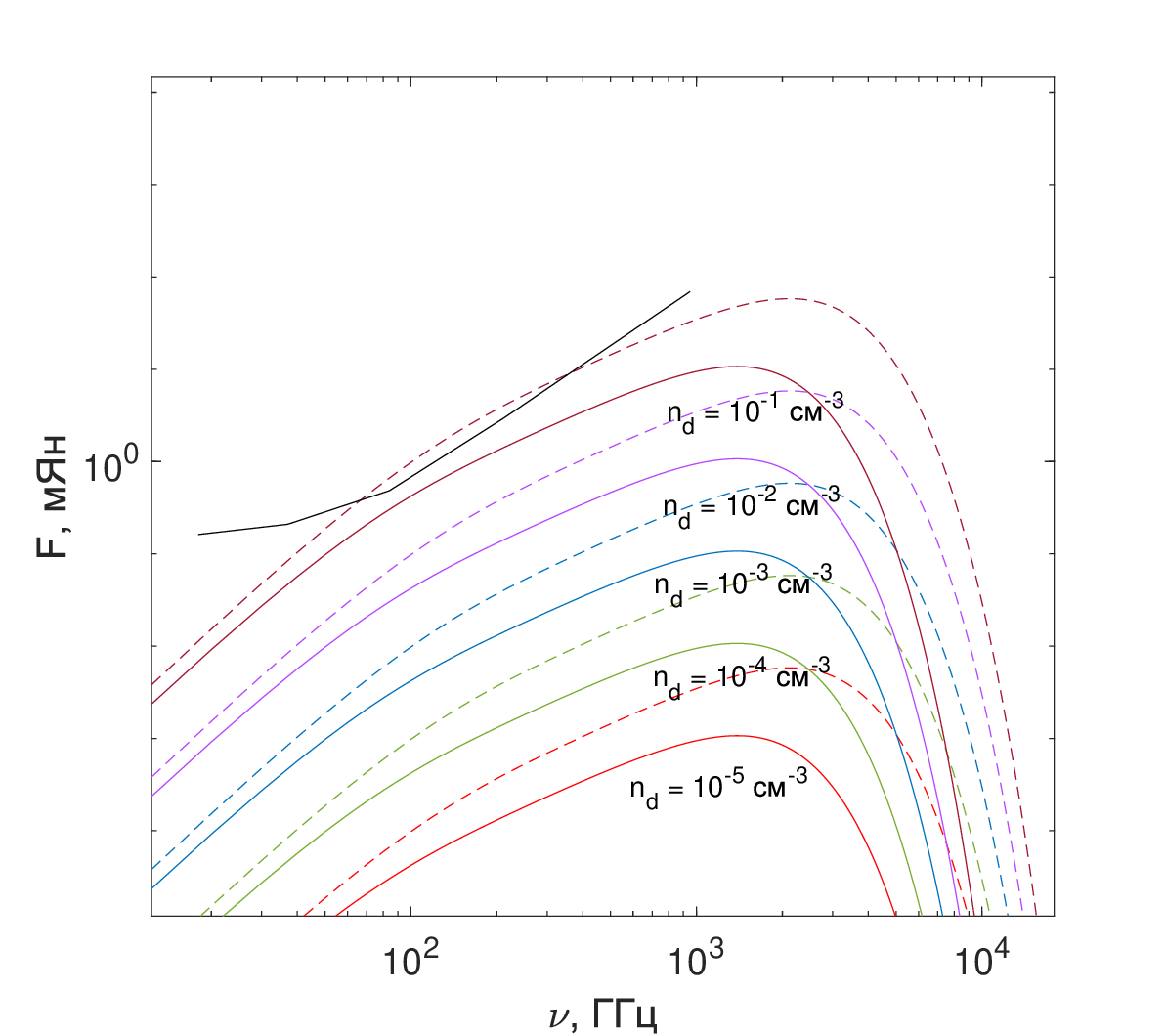}
\includegraphics[width=0.49\textwidth]{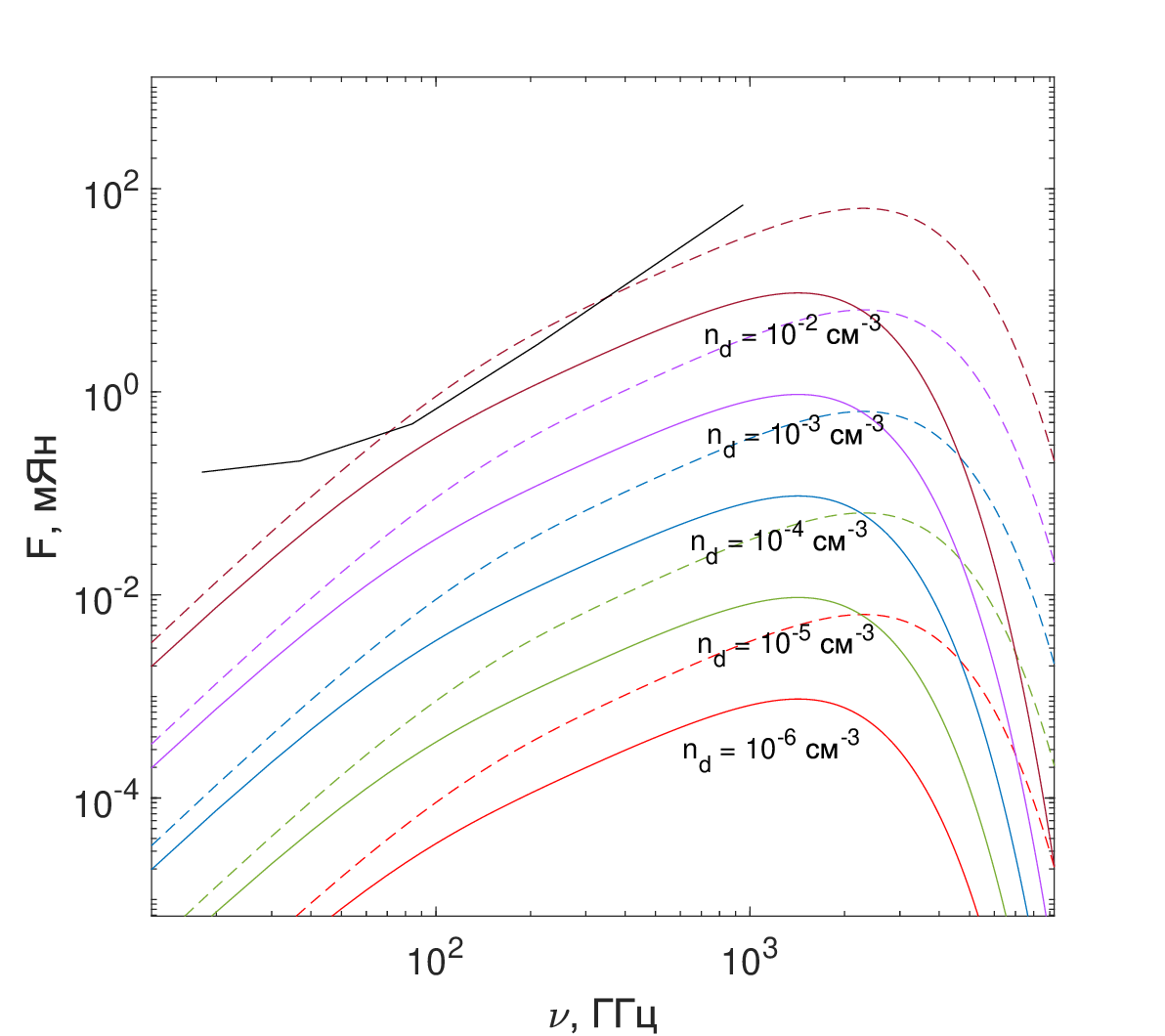}\\
\includegraphics[width=0.49\textwidth]{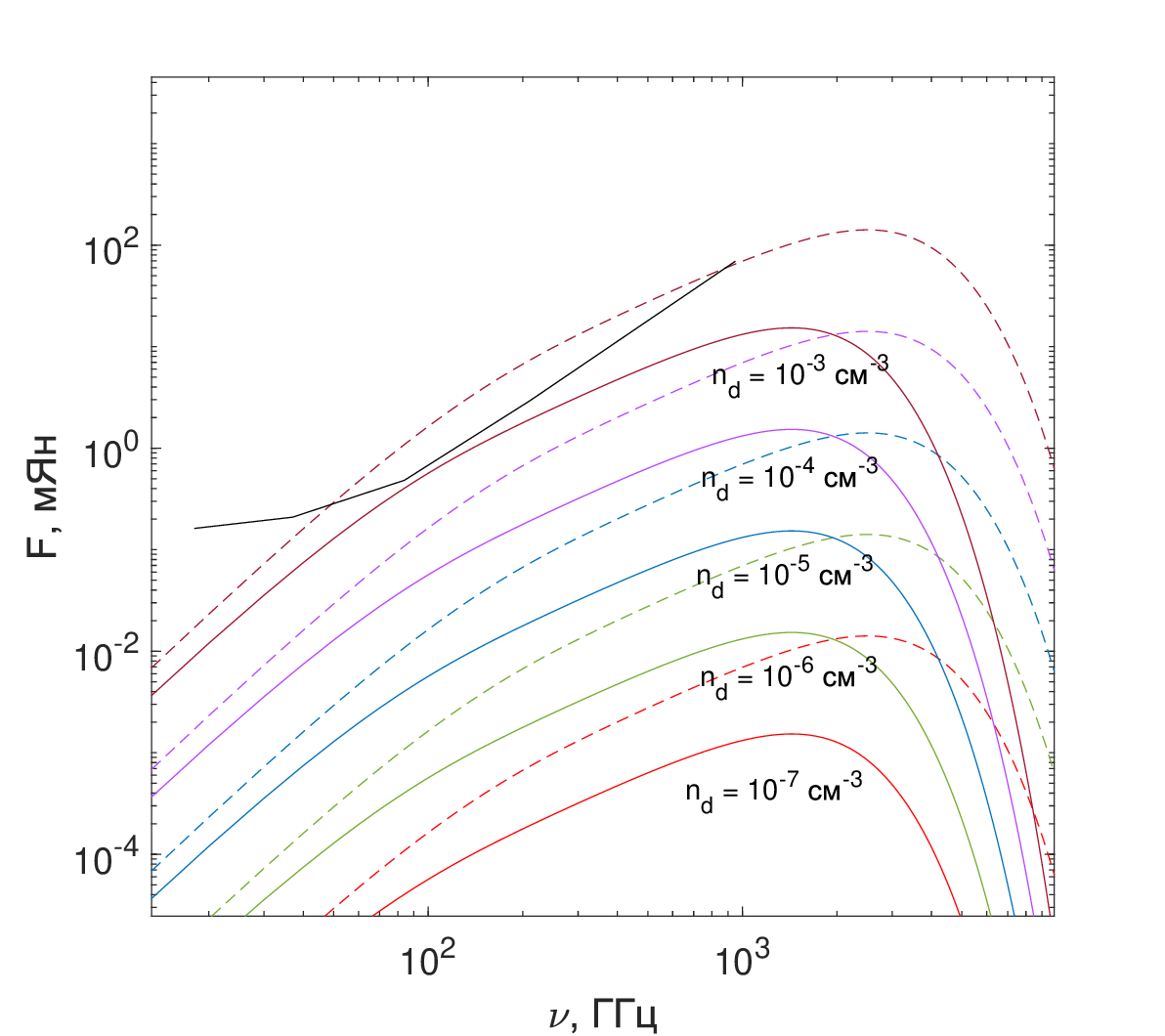}
\includegraphics[width=0.49\textwidth]{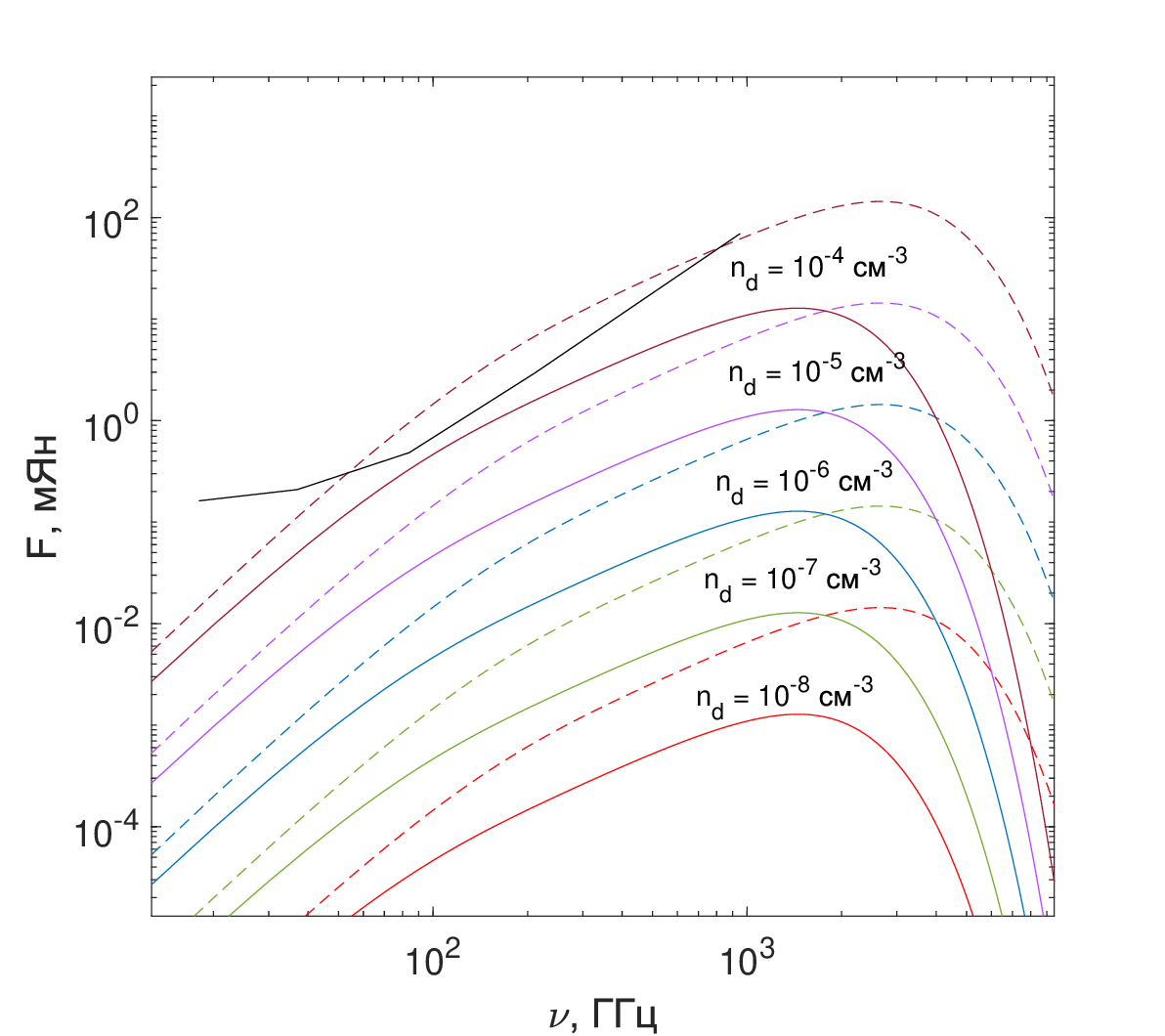}
\caption{Same as in Fig.~5, but for $M=10^{18}$~g.}
\end{figure*}
%\twocolumn

In work \cite{jefferson052018} this model was applied to the CMZ of the Galaxy. As a result, the lifetime of clouds at a distance $\sim 45-120$~pc from the Galactic center ranges from 1.4 to 3.9 million years. In this case, the evolution of interstellar clouds is dominated by the influence of gravitational collapse. At other scales of the CMZ, the contribution of galactic shear dominates, with the lifetime of interstellar clouds ranging from 3 to 9 million years. Thus, the lifetime of a molecular cloud is much shorter than the lifetime of PBHs of the masses considered in this work. 

%==============================fig8
%\onecolumn
\begin{figure*}
%\label{fig:spec_10_19}
\includegraphics[width=0.49\textwidth]{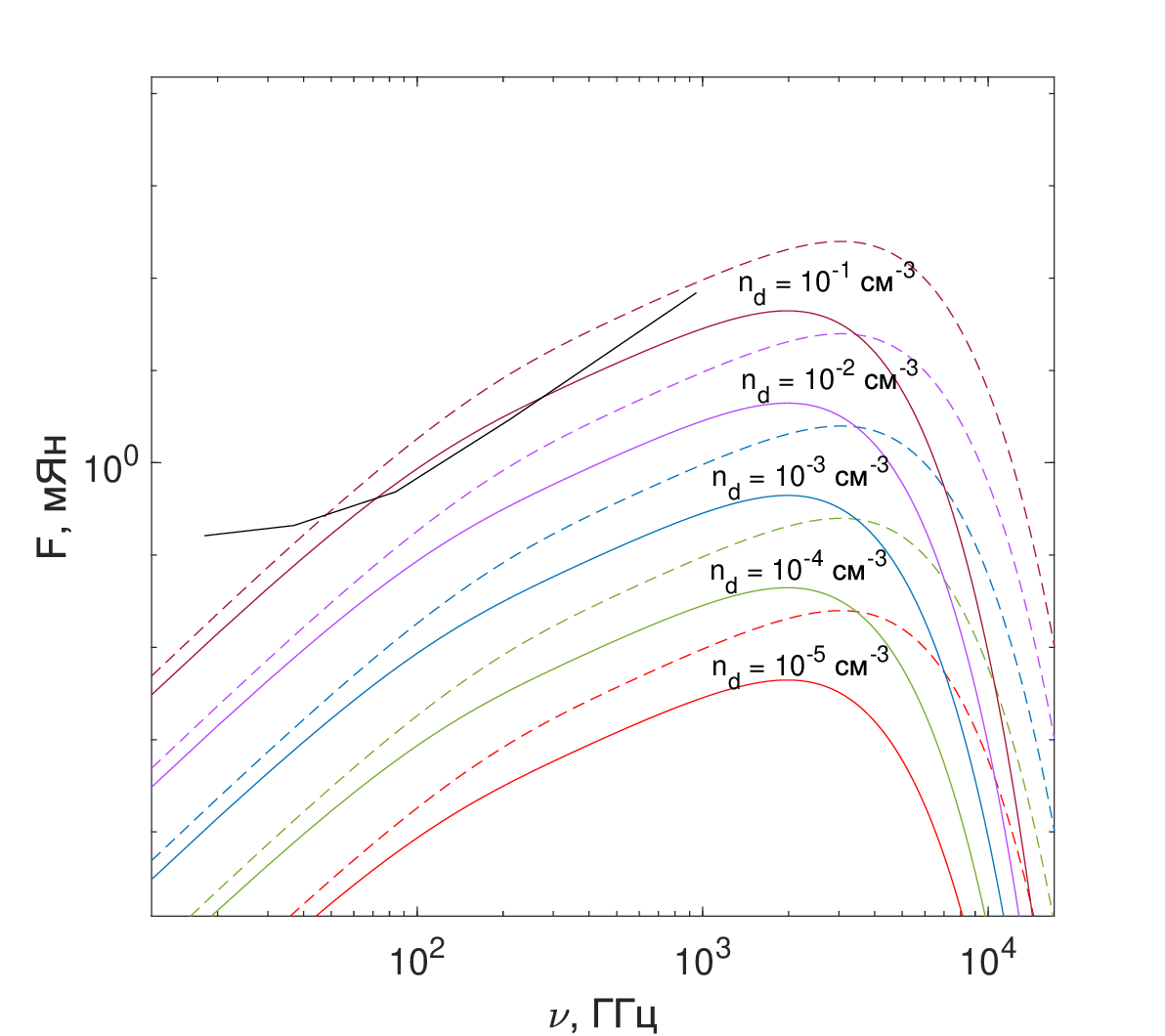}
\includegraphics[width=0.49\textwidth]{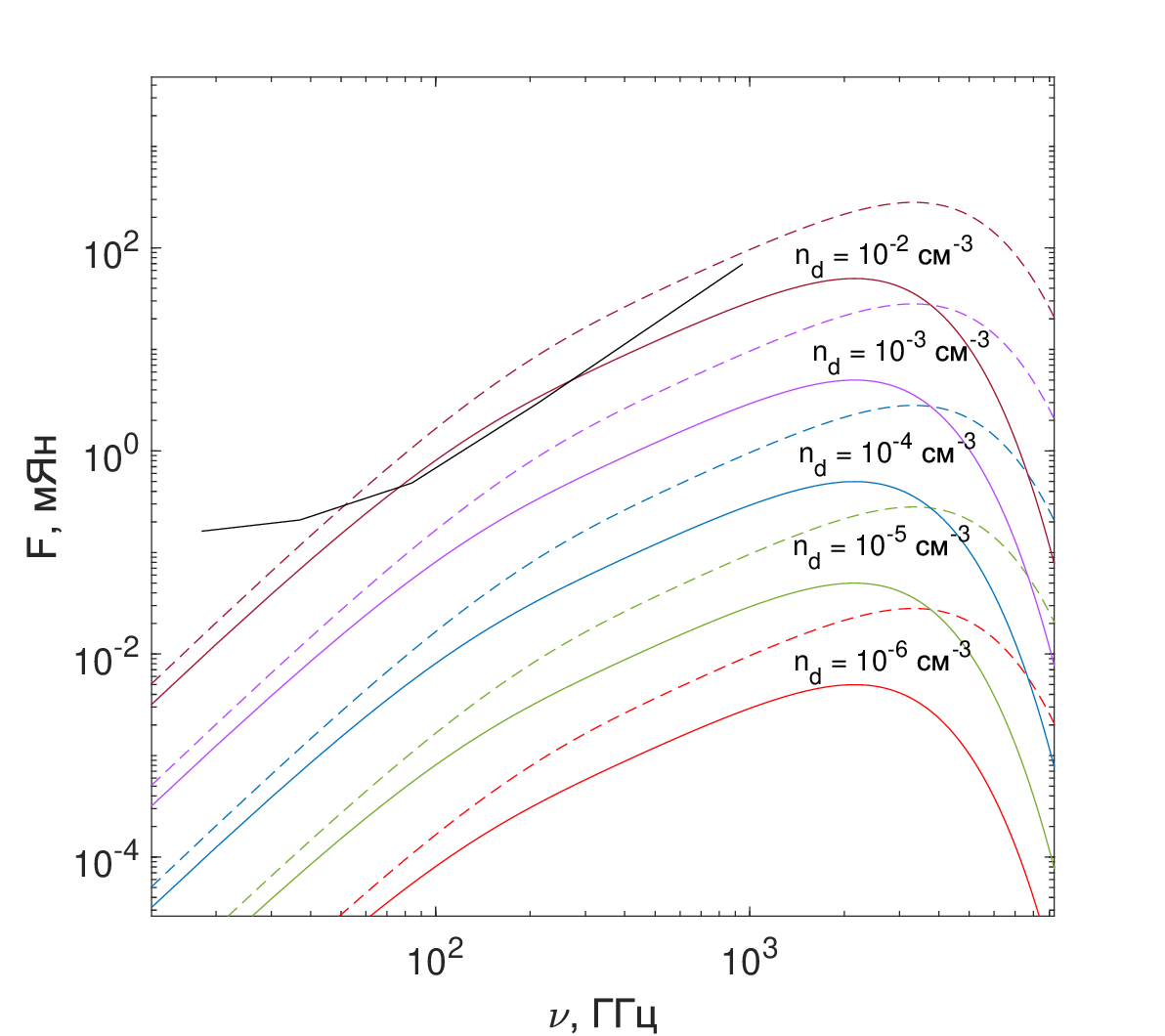}\\
\includegraphics[width=0.49\textwidth]{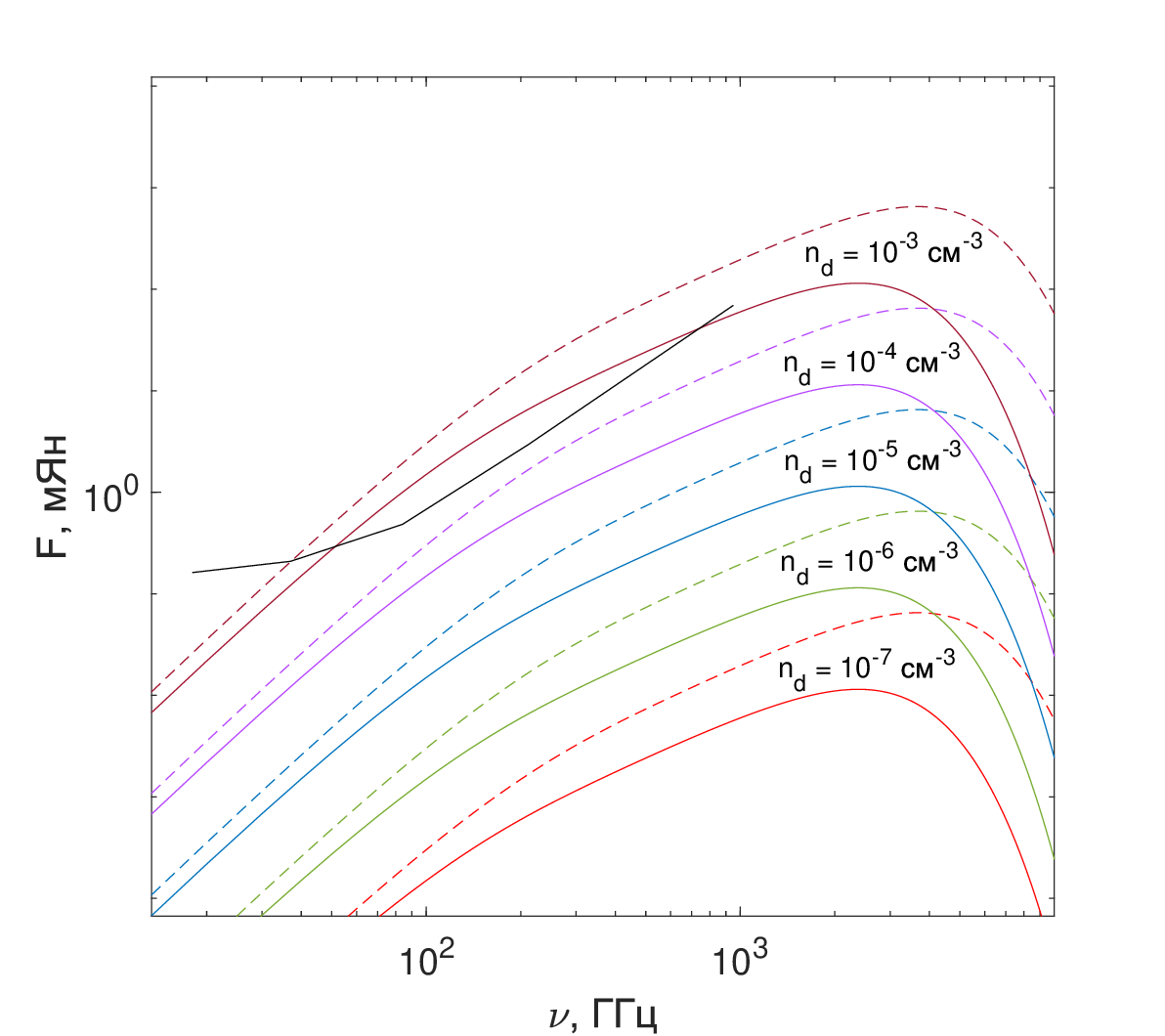}
\includegraphics[width=0.49\textwidth]{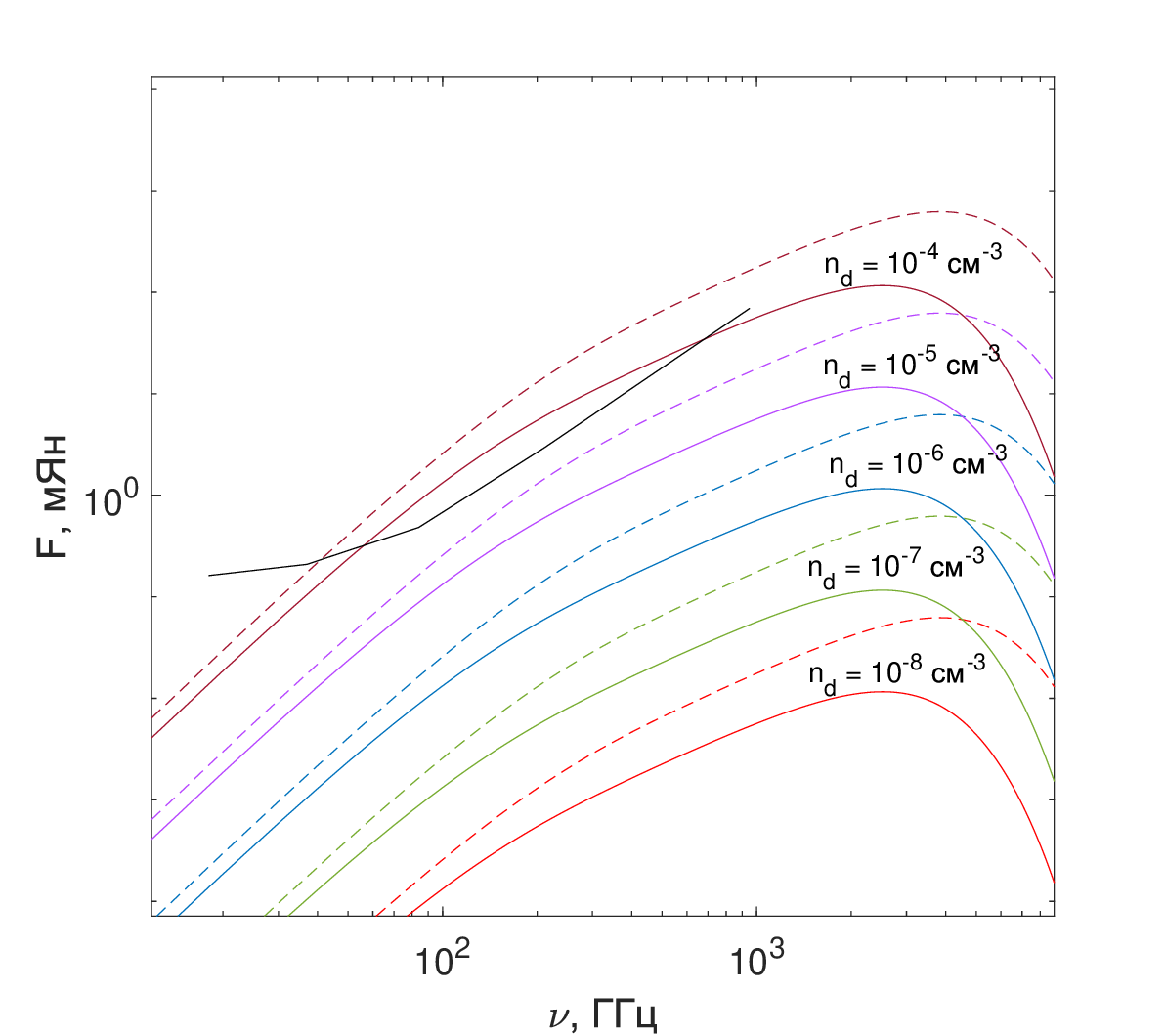}
\caption{Same as in Fig.~5, but for $M=10^{19}$~g.}
\end{figure*}
%\twocolumn

%=================
\section{CONCLUSIONS}
%=================
The paper examines the process of dust heating by photons from a PBH located in a molecular cloud
with masses of $M=10^{16} - 10^{20}$~g. Assuming that dust grains are uniformly distributed in a spherically symmetric cloud and have sizes $a = 0.01, 0.02, 0.05$, and $0.1~\mu\rm m$ the dust temperature was calculated as a function of distance from the PBH. The obtained graphs show that directly next to the PBH, dust grains are heated to temperatures of $T \lesssim 10^2$~K, and with increasing distance, the temperature drops sharply. Thus, the PBH can heat only a spherical
layer of the molecular cloud with a radius $r$ of no more than $10^{3}$~cm. Additionally, in this paper, emission spectra of dust grains heated by PBH were constructed and it is shown how the spectra depend on dust concentration in the molecular cloud and dust grain sizes. The obtained spectra were added with sensitivity graphs of the planned Millimetron Space Observatory in interferometer mode, and it was shown that if a PBH is present in the cloud, it can be detected if the dust concentration in the cloud ranges from $n_d=10^{-4}$~cm$^{-3}$ to $n_d=10^{-1}$~cm$^{-3}$ (while the size of individual dust grains in the cloud varies from 0.1 to 0.01~$\mu\rm m$, respectively).

%============================fig9
%\onecolumn
\begin{figure*}
%\label{fig:spec_10_20}
\includegraphics[width=0.49\textwidth]{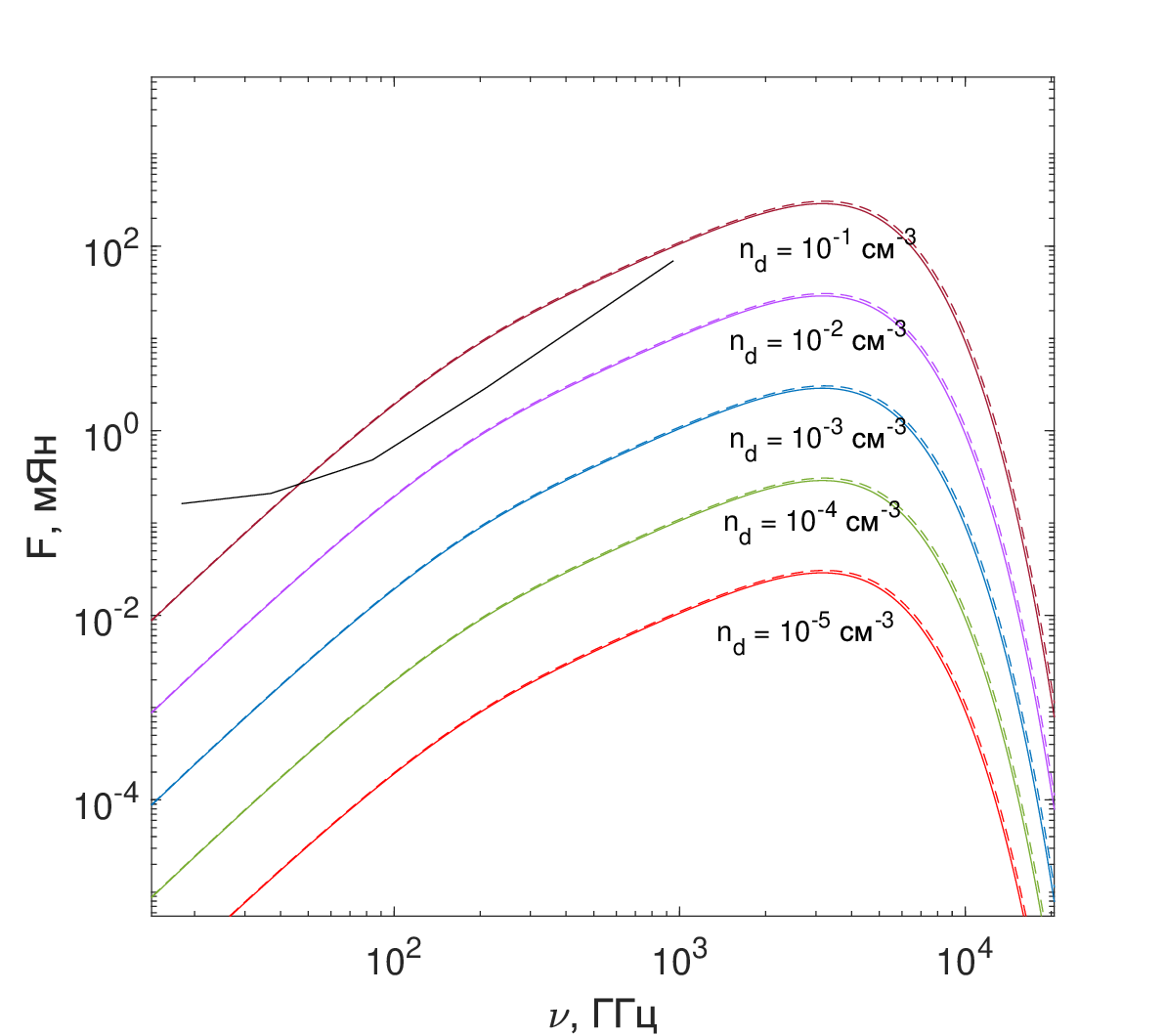}
\includegraphics[width=0.49\textwidth]{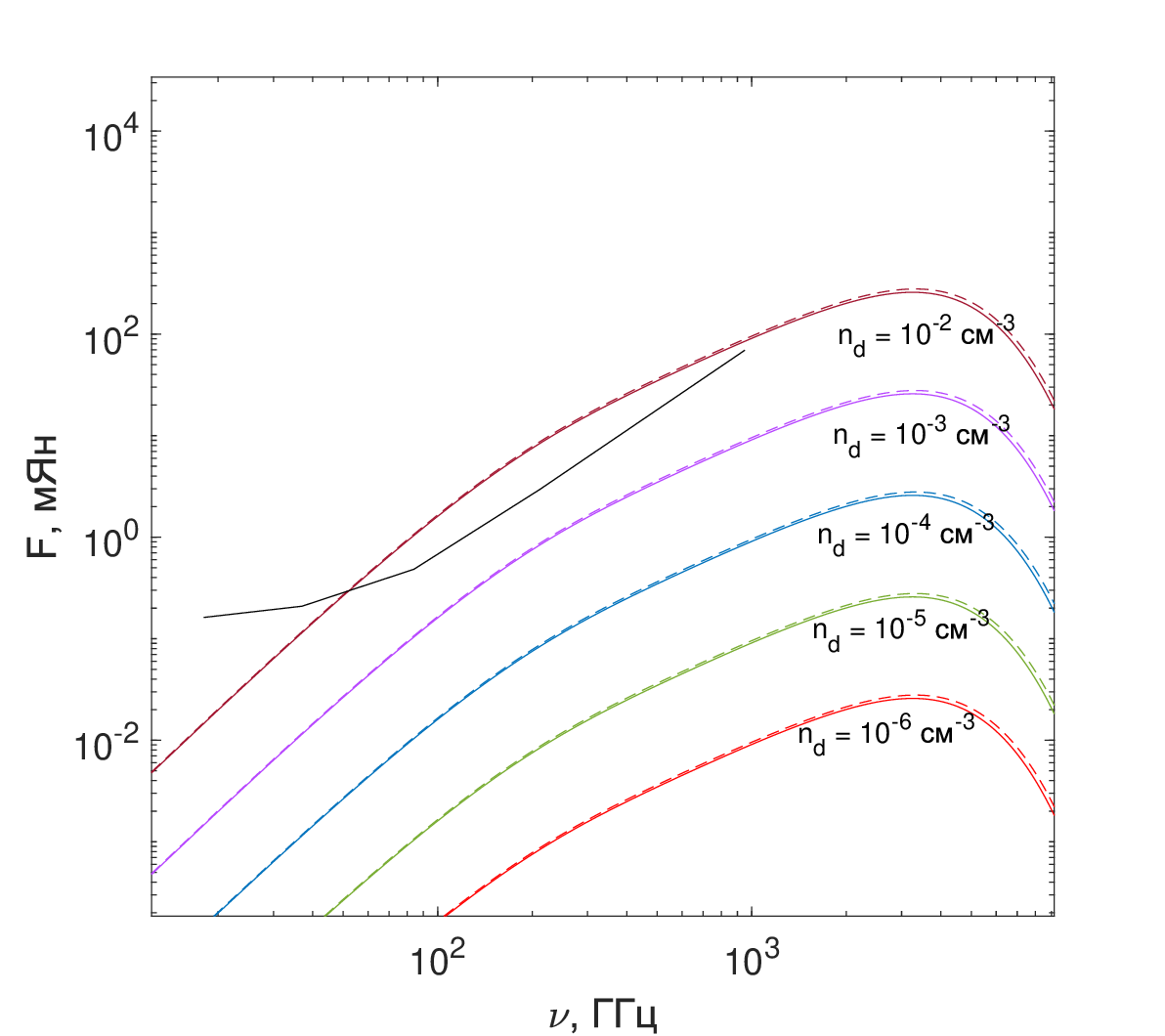}\\
\includegraphics[width=0.49\textwidth]{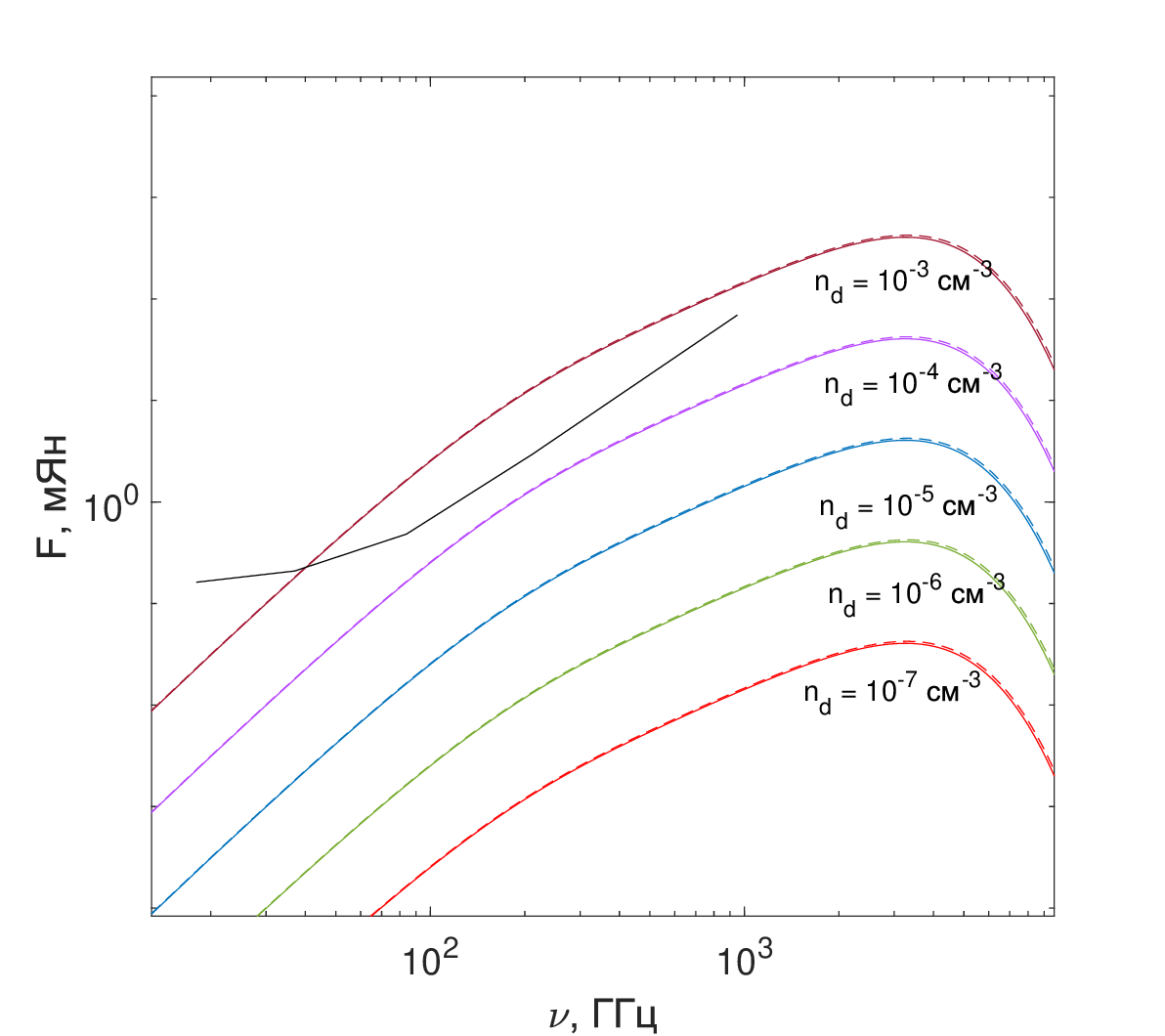}
\includegraphics[width=0.49\textwidth]{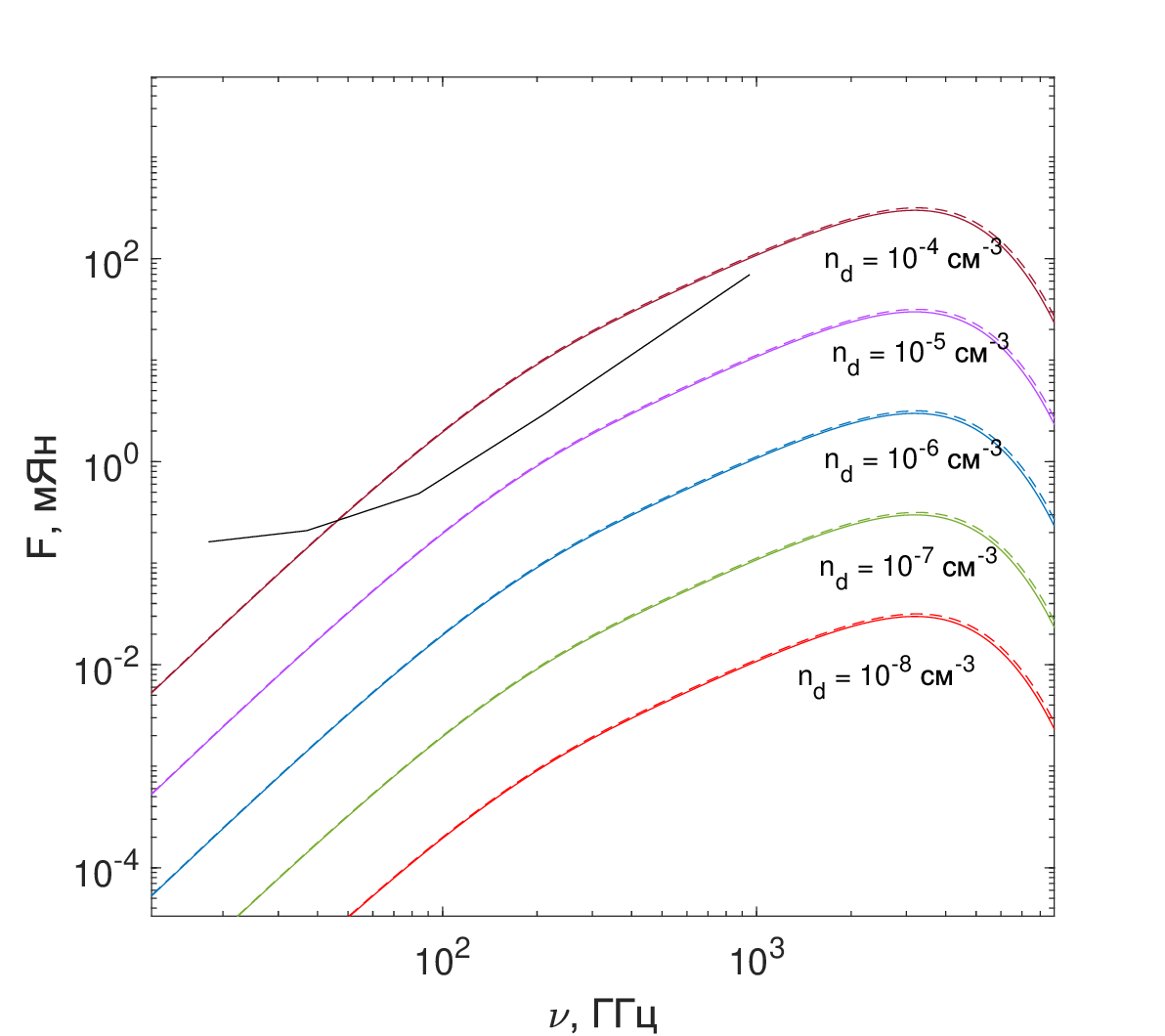}
\caption{Same as in Fig.~5, but for $M=10^{20}$~g.}
\end{figure*}
%\twocolumn
%================================================================

%%%%%%%%%%%%%%%%%
\section*{ACKNOWLEDGMENTS}
%%%%%%%%%%%%%%%%%

The authors are grateful to the reviewer, whose questions helped to clarify certain aspects of the
article.

%%%%%%%%%%
\section*{FUNDING}
%%%%%%%%%%
This work was supported by the Russian Science Foundation (grant No. 24-22-00146).

%============

\end{document}